\documentclass[authoryear,preprint,review,12pt]{elsarticle}

\usepackage{amssymb}

\usepackage{amsmath}
\usepackage{amsfonts}
\usepackage{subfigure}
\usepackage{epstopdf}

\journal{New Astronomy Reviews}

\begin{document}

\begin{frontmatter}



\title{Stellar Intensity Interferometry: \\
    Prospects for sub-milliarcsecond optical imaging}


\author[a]{Dainis Dravins}
\ead{dainis@astro.lu.se}
\author[b]{Stephan LeBohec}
\author[a,c]{Hannes Jensen}
\author[b]{\& Paul D. Nu{\~n}ez}

\address[a]{Lund Observatory, Box 43, SE-22100 Lund, Sweden}
\address[b]{Department of Physics and Astronomy, The University of Utah, 115 South 1400 East, Salt Lake City, UT 84112-0830, U.S.A.}
\address[c]{Present address: Department of astronomy, Stockholm University,\\ AlbaNova University Center, SE-10691 Stockholm, Sweden}


\begin{abstract}
Using kilometric arrays of air Cherenkov telescopes at short wavelengths, intensity interferometry may increase the spatial resolution achieved in optical astronomy by an order of magnitude, enabling images of rapidly rotating hot stars with structures in their circumstellar disks and winds, or mapping out patterns of nonradial pulsations across stellar surfaces.  Intensity interferometry (once pioneered by Hanbury Brown and Twiss) connects telescopes only electronically, and is practically insensitive to atmospheric turbulence and optical imperfections, permitting observations over long baselines and through large airmasses, also at short optical wavelengths.  The required large telescopes ($\sim$\,10 m) with very fast detectors ($\sim$\,ns) are becoming available as the arrays primarily erected to measure Cherenkov light emitted in air by particle cascades initiated by energetic gamma rays.  Planned facilities (e.g., CTA, {\it{Cherenkov Telescope Array}}) envision many tens of telescopes distributed over a few square km. Digital signal handling enables very many baselines (from tens of meters to over a kilometer) to be simultaneously synthesized between many pairs of telescopes, while stars may be tracked across the sky with electronic time delays, in effect synthesizing an optical interferometer in software.  Simulated observations indicate limiting magnitudes around m$_{V}$=\,8, reaching angular resolutions $\sim$30$\,{\mu}$arcsec in the violet. The signal-to-noise ratio favors high-temperature sources and emission-line structures, and is independent of the optical passband, be it a single spectral line or the broad spectral continuum.  Intensity interferometry directly provides the modulus (but not phase) of any spatial frequency component of the source image; for this reason a full image reconstruction requires phase retrieval techniques.  This is feasible if sufficient coverage of the interferometric $(u,v)-$plane is available, as was verified through numerical simulations. Laboratory and field experiments are in progress; test telescopes have been erected, intensity interferometry has been achieved in the laboratory, and first full-scale tests of connecting large Cherenkov telescopes have been carried out.  This paper reviews this interferometric method in view of the new possibilities offered by arrays of air Cherenkov telescopes, and outlines observational programs that should become realistic already in the rather near future.

\end{abstract}


\end{frontmatter}



\section{Highest resolution in astronomy}

Much of astronomy is driven by imaging with improved spatial resolution and science cases for constantly higher resolution are overwhelming.  Our local Universe is teeming with stars but astronomers are still basically incapable of observing stars as such.  We do observe the light radiated by them but -- with few exceptions -- are still unable to observe the stars themselves, i.e., resolve their disks or view structures across and outside their surfaces (except for the Sun, of course).  One can just speculate what new worlds will be revealed once stars will no longer be seen as mere point sources but as extended and irregular objects with magnetic or thermal spots, flattened or distorted by rapid rotation, and with mass ejections through their circumstellar shells monitored in different spectral features as they flow towards their binary companions.  It is not long ago that the satellites of the outer planets passed from being mere point sources to a plethora of different worlds, and one could speculate what meager state extragalactic astronomy would be in, were galaxies observed as point sources only.

Tantalizing results from current optical interferometers show how stars are beginning to be seen as a vast diversity of objects, and a great leap forward will be enabled by improving angular resolution by just another order of magnitude.  Bright stars have typical diameters of a few milliarcseconds, requiring optical interferometry over hundreds of meters or some kilometer to enable surface imaging. However, amplitude (phase-) interferometers require optical precisions of both their optics, and of the atmosphere above, to within a small fraction of a wavelength, and atmospheric turbulence constrains their operation when baselines exceed some 100 m, especially at shorter visual wavelengths.  Using a simple {$\lambda$}/$r$ criterion for the required optical baseline, a resolution of 1 milliarcsecond (mas) at {$\lambda$}~500 nm requires a length around 100 meters, while 1 km enables 100 ${\mu}$as.
 
The potential of very long baseline optical interferometry for imaging stellar surfaces has been realized by several (e.g., Labeyrie 1996; Quirrenbach 2004), and proposed concepts include extended amplitude interferometer arrays in space: {\it{Stellar Imager}} (Carpenter et al.\ 2007) and the {\it{Luciola hypertelescope}} (Labeyrie et al.\ 2009), or possibly placed at high-altitude locations in Antarctica (Vakili et al.\ 2005).  However, despite their scientific appeal, the complexity and probable expense of these projects make the timescales for their realization somewhat uncertain, prompting searches for alternative approaches.  One promising possibility is ground-based intensity interferometry.

\subsection{Intensity interferometry} 

Intensity interferometry was pioneered by Robert Hanbury Brown and Richard Q.\ Twiss already long ago (Hanbury Brown 1974) for the original purpose of measuring stellar sizes, and a dedicated instrument was built at Narrabri, Australia.  What is observed is the second-order coherence of light (i.e., that of intensity, not of amplitude or phase), by measuring temporal correlations of arrival times between photons recorded in different telescopes. At the time of its design, the understanding of its functioning was a source of considerable confusion (although it was explained in terms of classical optical waves undergoing random phase shifts), and even now it may be challenging to intuitively comprehend.  Somewhat later, a more complete semi-classical theory was developed (e.g., 
Mandel \& Wolf 1995).  Seen in a quantum context, this is a two-photon process, and the intensity interferometer is often seen as the first quantum-optical experiment.  It laid the foundation for a series of experiments of photon correlations including also states of light that do not have classical counterparts (such as photon antibunching).  A key person in developing the quantum theory of optical coherence was Roy Glauber (1963abc; 2007), acknowledged with the 2005 Nobel prize in physics.

The name {\it{intensity interferometer}} itself is sort of a misnomer: there actually is nothing interfering in the instrument, rather its name was chosen for its analogy to the ordinary amplitude interferometer, which at that time had similar scientific aims in measuring source diameters.  Two separate telescopes are simultaneously measuring the random and very rapid intrinsic fluctuations in the light from some particular star.  When the telescopes are placed sufficiently close to one another, the fluctuations measured in both telescopes are correlated, but when moving them apart, the fluctuations gradually become decorrelated.  How rapidly this occurs for increasing telescope separations gives a measure of the spatial coherence of starlight, and thus the spatial properties of the star.  The signal is a measure of the second-order spatial coherence, the square of that visibility which would be observed in any classical amplitude interferometer, and the spatial baselines for obtaining any given resolution are thus the same as would be required in ordinary interferometry.

The great observational advantage of intensity interferometry (compared to amplitude interferometry) is that it is practically insensitive to either atmospheric turbulence or to telescope optical imperfections, enabling very long baselines as well as observing at short optical wavelengths, even through large airmasses far away from zenith.  Telescopes are connected only electronically (rather than optically), from which it follows that the noise budget relates to the relatively long electronic timescales (nanoseconds, and light-travel distances of centimeters or meters) rather than those of the light wave itself (femtoseconds and nanometers).  A realistic time resolution of perhaps 10 nanoseconds corresponds to 3\,m light-travel distance, and the control of atmospheric path-lengths and telescope imperfections then only needs to correspond to some reasonable fraction of those 3 meters. 

Since the measured quantity is the {\it{square}} of the ordinary visibility, it always remains positive (save for measurement noise), only diminishing in magnitude when smeared over time intervals longer than the optical coherence time of starlight (due to finite time resolution in the electronics or imprecise telescope placements along the wavefront).  However, for realistic time resolutions (much longer than an optical coherence time of perhaps $\sim$10$^{-14}$~s), the magnitude of any measured signal is tiny, requiring very good photon statistics for its reliable determination.  Large photon fluxes (and thus large telescopes) are therefore required; already the flux collectors used in the original intensity interferometer at Narrabri, were larger than any other optical telescope at that time.  Although the signal can be enhanced by improving the electronic time resolution, faster electronics can only be exploited up to a point since there follows a matching requirement on the optomechanical systems.  A timing improvement to 100 ps, say, would require mechanical accuracies on mm levels, going beyond what typically is achieved in flux collectors, and beginning to approach the level of fluctuations in path-length differences induced by atmospheric turbulence (Cavazzani et al.\ 2012; Wijaya \& Brunner 2011).

Details of the original intensity interferometer at Narrabri and its observing program were documented in Hanbury Brown et al.\ (1967ab), with retrospective overviews by Hanbury Brown (1974; 1985; 1991).  The principles are also explained in various textbooks and reference publications, e.g., Glindemann (2011), Goodman (1985), Loudon (2000), Mandel \& Wolf (1995), and very lucidly in Labeyrie et al.\ (2006), Saha (2011), and Shih (2011).

The original intensity interferometer at Narrabri had two reflecting telescopes of 6.5\,m diameter, formed by mosaics of 252 hexagonal mirrors, providing images of 12\,arcmin diameter.  In order to maintain a fixed baseline while tracking (and to avoid the need for variable signal delays), the telescopes moved on a railway track of 188\,m diameter.  (The design parameters are said to have been chosen to enable it to spatially resolve the O5 star {$\zeta$}~Puppis).  Its main observing program, completed in 1973, measured angular diameters of 32 stars brighter than about m$_{V}$=\,2.5 and hotter than T$_{eff}$= 7000\,K,  producing an effective-temperature scale for early-type stars of spectral types between O5 and F8.  Following the completion of that program, the design for a second-generation intensity interferometer was worked out (Davis 1975; Hanbury Brown 1979; 1991), envisioned to have 12-m diameter telescopes, movable over 2\,km.  However, the then concurrent developments in astronomical amplitude interferometry, demonstrated already with very small telescopes, led this Australian group in that direction instead and (although a few experiments have been made in radio; e.g., Erukhimov et al.\ 1970) astronomical intensity interferometry saw no further development.

However, following its start in astronomy, intensity interferometry has been vigorously pursued in other fields, both for studying optical light in the laboratory, and in analyzing interactions in high-energy particle physics. For laboratory studies of scattered light, photon correlation spectroscopy can be considered as intensity interferometry in the temporal (not spatial) domain, and is a tool to measure the {\it{temporal}} coherence of light, and to deduce its spectral broadening (e.g.,  Becker 2005; Degiorgio \& Lastovka 1971; Oliver 1978; Saleh 1978).

In particle physics, the same basic quantum principles of measuring intensity correlations apply to all bosons, i.e., particles which -- just like photons -- carry an integer value of quantum spin, and therefore share the same type of Bose-Einstein quantum statistics (Alexander 2003; Baym 1998; Boal et al.\ 1990).  In a 1959 bubble-chamber study of charged pion production in proton/antiproton annihilation, the angular distribution of like-charge pion-pairs was found to differ from the unlike-charge ones.  In a now classic paper (Goldhaber et al.\ 1960), this was interpreted as due to Bose-Einstein correlations, although the realization that the effect was equivalent to the astronomical intensity interferometer came only in the 1970's.  These studies in particle physics are now generally referred to as `HBT-interferometry' (for Hanbury Brown-Twiss), although also terms such as `femtoscopy' or just `Bose-Einstein correlations' are used.  A historical overview of that extensive field is by Padula (2005), while more current activities are reviewed by Bauer et al.\ (1992); Cs{\"o}rg{\H o} (2006); Heinz et al.\ (1999); Lisa et al.\ (2005); Wiedemann \& Heinz (1999), or in the monograph by Weiner (2000).

Thus, intensity interferometry has not been further pursued in astronomy since a long time ago, largely due to its demanding requirements for large and movable optical flux collectors, spread over long baselines, and equipped with fast detectors and high-speed electronics.  However, all of these requirements are now rapidly being satisfied through the combination of high-speed digital signal handling with the construction of telescope complexes, erected for a different primary purpose, namely to optically record atmospheric Cherenkov light for the study of the most energetic gamma rays.

The purpose of this paper is to review this interferometric method in view of such new possibilities, and to outline observational programs that should become realistic already in the rather near future.  Current efforts to develop the several stages required towards its realization are described, including several issues that are specific to this method.  Given that no astronomical intensity interferometer currently exists -- and that its functioning is fundamentally different from that of any other astronomical imaging system -- this review aims at connecting the past pioneering efforts by Hanbury Brown et al., with the potential offered by the forthcoming new arrays of air Cherenkov telescopes, and to explain the possibilities (and limitations) also for readers that might not yet be familiar with these techniques.

\begin{figure}
\includegraphics[width=9cm, angle=90]{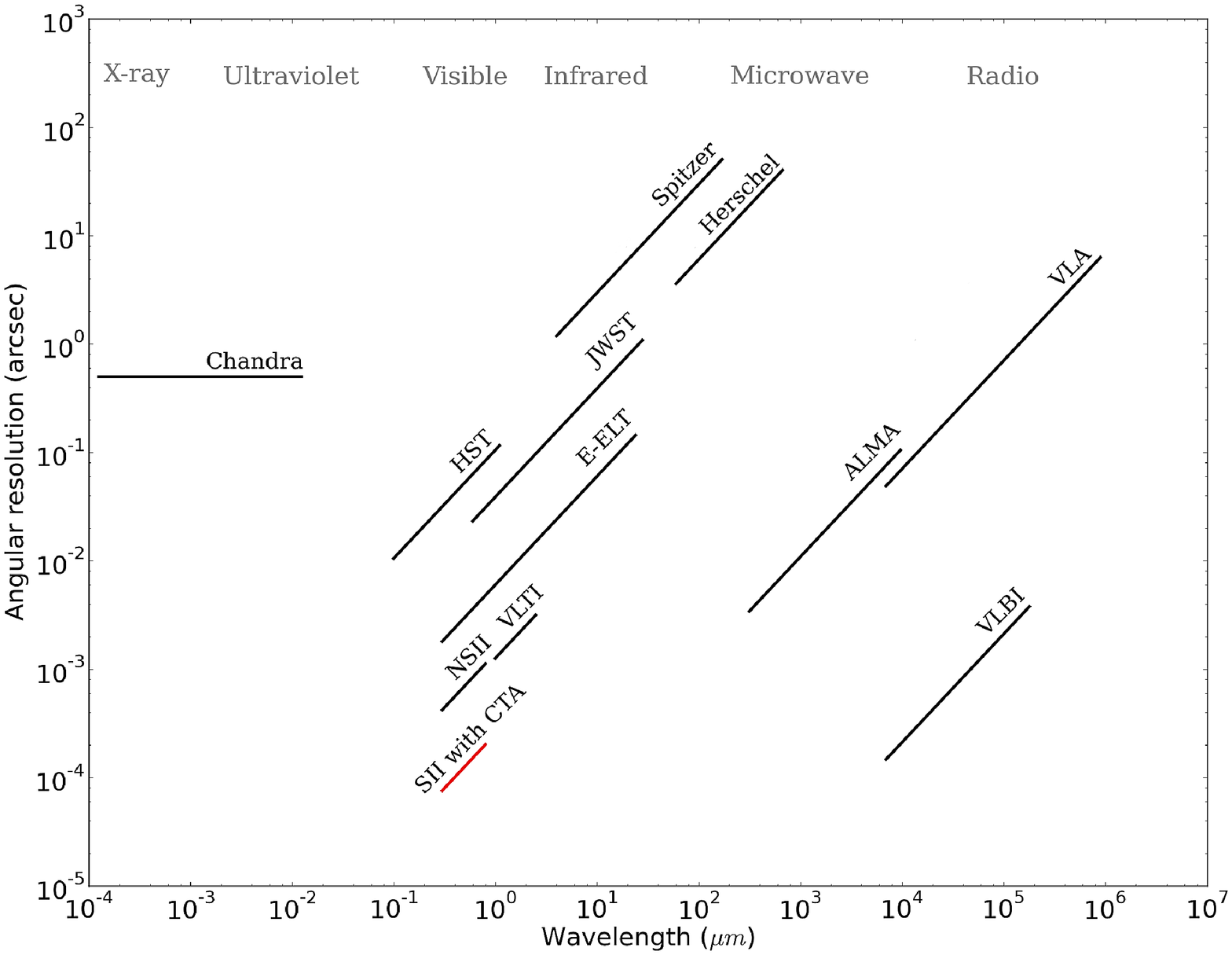}
\centering
\caption{Angular resolution for existing and future observatories at different wavelengths.  Except for X-rays, resolutions were taken as diffraction-limited.  HST = Hubble Space Telescope; JWST = James Webb Space Telescope; NSII = Narrabri Stellar Intensity Interferometer; E-ELT = European Extremely Large Telescope; VLTI = Very Large Telescope Interferometer; VLA = Very Large Array; ALMA = Atacama Large Millimeter Array; VLBI = Very Long Baseline Interferometry (here for a baseline equal to the Earth diameter); CTA = Cherenkov Telescope Array.  Intensity interferometry with large Cherenkov arrays offers unprecedented angular resolution, challenged only by radio interferometers operating between Earth and antennas in deep space. \label{fig1}}
\end{figure}

\begin{figure}
\centering
\includegraphics[width=6cm, angle=90]{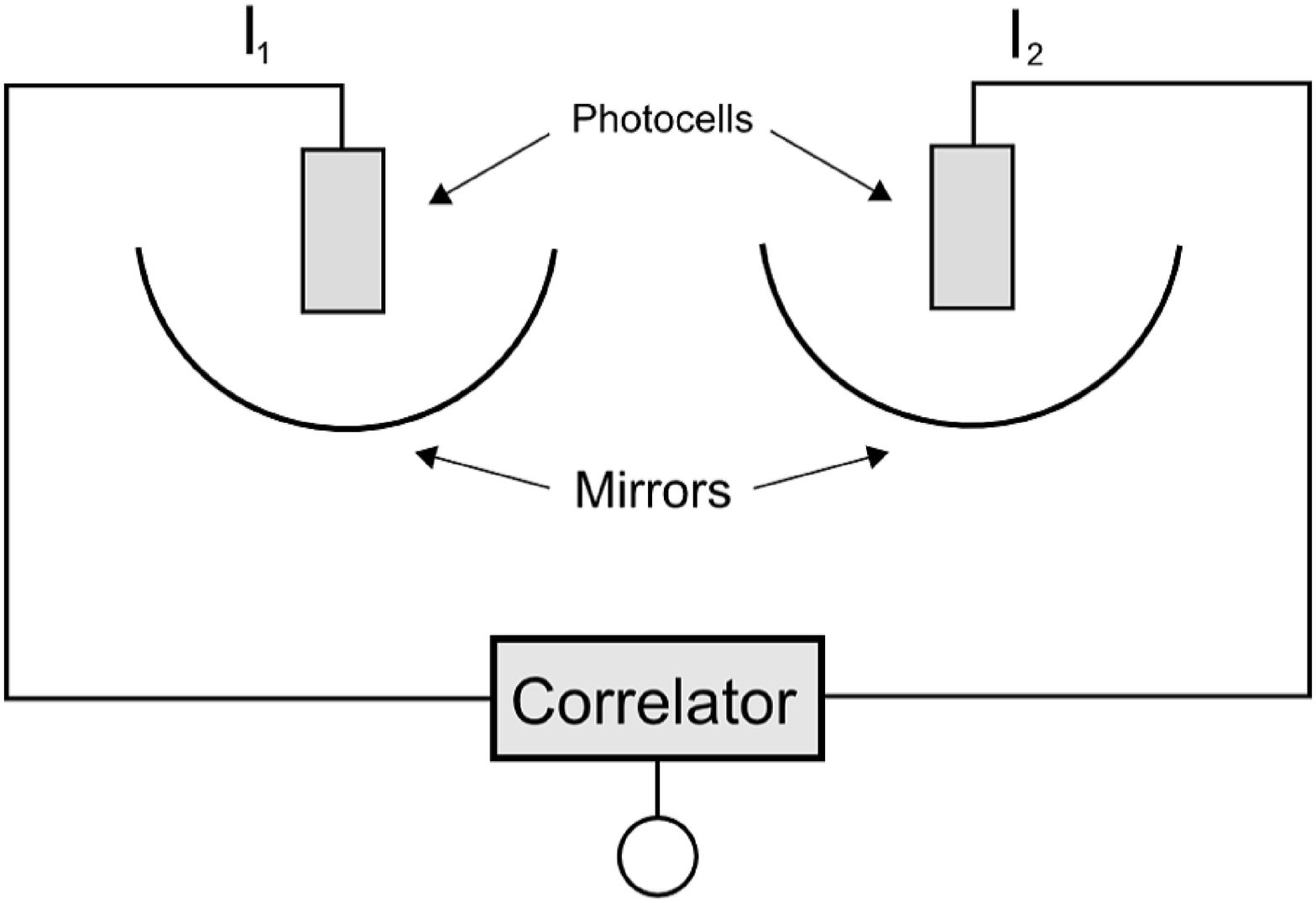}
\caption{Basic components of an intensity interferometer.  Two telescopes observe the same source, and the measured time-variable intensities are electronically cross correlated.  \label{fig2}}
\end{figure}

\clearpage

\begin{table}[b]
\caption{Properties of the three examined configurations of Cherenkov telescope arrays (\#1 corresponds to the upper rows in Figures \ref{fig3} and \ref{magnitudes}; \#2 to the middle ones). $N$ is the number of telescopes, $A$ is the light collection area of each type of telescope, $b$ is the number of unique baselines available, $B_{min}, B_{max}$ indicates the range of baselines for observations in zenith. The corresponding range of angular diameters in milliarcseconds $(1.22 \lambda/r)$ for observations at $\lambda$~400~nm is indicated by  $\theta_{min}, \theta_{max}$.   In the original CTA design study, these three configurations were designated with the letters B, D, and I (Actis et al.\ 2011).}
\begin{center}
\begin{tabular}{lrrrrr|}
Array & $N$ & $A$ [m$^2$] & $b$ & $B_{min}, B_{max}$ [m] & $\theta_{min}, \theta_{max}$ [mas]\\
\hline \hline
\#1 & 42 & 113, 415 & 253 & 32, 759 & 0.13, 3.2 \\
\#2 & 57 & 113 & 487 & 170, 2180 & 0.05, 0.6 \\
\#3 & 77 & 28, 113, 415 & 1606 & 90, 2200 & 0.05,  1.13	
\end{tabular}
\end{center}
\label{configurationstable}
\end{table}

\subsection{Air Cherenkov telescopes} 

Seemingly ideal flux collectors for intensity interferometry are those air Cherenkov telescopes that are being erected for gamma-ray astronomy.  These measure the feeble and brief flashes of Cherenkov light produced in air by cascades of secondary particles initiated by very energetic gamma rays.  Time resolution has to be no worse than a few nanoseconds (duration of the Cherenkov light flash); they must be sensitive to short optical wavelengths (Cherenkov light is bluish); they must be large (Cherenkov light is faint), and they must be spread out over hundreds of meters (size of the Cherenkov light-pool onto the ground).  The image seen by any one telescope shows the track of the air shower, but multiple telescopes are required for a more precise stereoscopic reconstruction of the shower geometry, and thus the direction to the source. For imaging the air shower, a modest optical imaging quality is sufficient (3--5 arcminutes, say), but possibly diverse path-lengths within the optics must not temporally smear out the Cherenkov pulse more than a few nanoseconds.  The success of this concept has prompted the recent construction of several arrays with large flux collectors, including H.E.S.S. in Namibia, MAGIC on La Palma, and VERITAS in Arizona.  These telescopes are large (12\,m diameter for VERITAS, 17\,m for MAGIC, and for H.E.S.S. even one 28\,m dish is being completed), distributed over distances of typically 50-200 meters (V{\"o}lk \& Bernl{\"o}hr 2009).  

These telescope parameters are remarkably similar to the requirements for intensity interferometry, and the compatibility is made even greater when realizing that the faintness of the Cherenkov light might preclude its efficient observation during brighter moonlight, a condition that does not inhibit interferometric observations of brighter sources (which can be made over a narrow optical bandwidth).  Further, electronic time delays can now be used to compensate for different arrival times of a wavefront to the different telescopes, removing the past requirement of having the telescopes continuously moving during observation.   

However, the most striking potential comes from planned future facilities which will improve both the gamma-ray flux sensitivity and the angular resolution by having great many, and widely distributed flux collectors.  The major international project is CTA, the Cherenkov Telescope Array (2012) which envisions a total of 50--100 telescopes with differently sized apertures between about 5 and 25 meters, distributed over an area of 2--3\,km{$^2$}.  Such an array permits an enormous number of baseline pairs to be synthesized, allowing to probe angular scales between milli- and microarcseconds.  The potential of using such arrays for intensity interferometry has indeed been noticed by several (e.g., de Wit et al.\ 2008; LeBohec \& Holder 2006; LeBohec et al.\ 2008a) and, within the CTA project, a working group now has the task to specify how to enable it for also intensity interferometry.  If a baseline of 2\,km could be utilized at {$\lambda$}\,350 nm, resolutions would approach 30\,{$\mu$}as.  This would offer unprecedented spatial resolution in astronomy (Figure \ref{fig1}), challenged only by radio interferometers operating between Earth and antennas in deep space (Kardashev 2009), or possibly future X-ray interferometers (MAXIM 2012).  In this paper, we will examine the methodology for those types of observations, and the astrophysical targets that may be imaged when entering the new microarcsecond parameter domain.

\section{Principles of intensity interferometry}

In its simplest form, an intensity interferometer consists of two telescopes or flux collectors, each with a photon detector feeding one channel of a correlator (Figure \ref{fig2}).

The intensities measured at detectors 1 and 2 are the respective values of the light-wave amplitude times its complex conjugate, averaged over some time interval corresponding to the signal bandwidth of the detectors and associated electronics:

\begin{equation}
\langle I(t) \rangle = \langle E(t)E^*(t) \rangle
\end{equation}

where $^*$ marks complex conjugate and $\langle~\rangle$ denotes averaging over time.  The intensities  measured in the two telescopes are cross correlated:

\begin{equation}
 \langle I_1(t) I_2(t) \rangle = \langle E_1(t)E^*_1(t) \cdot E_2(t)E^*_2(t) \rangle
\end{equation}

This expression can be expanded by dividing the complex field amplitudes into their real and imaginary parts.  Here one must make an assumption that is fundamental to the operation of an intensity interferometer: the light must be chaotic, i.e., with a Gaussian distribution of its temporally varying electric-field amplitudes; also called thermal- or maximum-entropy light (Bachor \& Ralph 2004; Foellmi 2009; Loudon 2000; Shih 2011).  However, there is no constraint on the optical passband or on the distribution of wavelengths of the light which may even be quasi-monochromatic, as long as the light waves undergo random phase shifts, so that an intensity fluctuation results over timescales equal to the optical coherence time.  For chaotic light, the real and imaginary parts of $E_1$ and $E_2$ are Gaussian random variates, i.e., the values of $E_1$ and $E_2$ measured at different times can be treated as random variables obeying a normal distribution.  Then the Gaussian moment theorem applies, which relates all higher-order correlations of Gaussian variates to products of their lower-order correlations (the mathematics of this is described in detail by Mandel \& Wolf 1995).{\footnote{One example of non-Gaussian light is that from an ideal laser, where the light wave undergoes no phase jumps, and where there are no intensity fluctuations in either time or space.  The nature of such light can be revealed from intensity correlations but an intensity interferometer cannot be used to deduce the spatial size of such a source.}}   It is then possible to show (e.g., Labeyrie 2006), that:

\begin{equation}
  \langle I_1(t) I_2 (t) \rangle = \langle I_1 \rangle \langle I_2 \rangle + |\Gamma_{12}|^2,
\end{equation}
 
or

 \begin{equation}
  \langle I_1(t) I_2 (t) \rangle = \langle I_1 \rangle \langle I_2 \rangle (1 + |\gamma_{12}|^2)
\end{equation}

where $|\Gamma_{12}|^2$, the second-order correlation function, corresponds to $\langle I_1 \rangle \langle I_2 \rangle |\gamma_{12}|^2$,  with $\gamma_{12}$ being the mutual coherence function of light between locations 1 and 2, the quantity commonly measured in amplitude interferometers.  This cross-correlation between $E_1$ and $E_2$ is:

\begin{equation}
\Gamma_{12}(\tau) = \langle E_1(t+\tau) E_2^*(t) \rangle.
\end{equation}

Or, more explicitly, for an integration time $T$:

\begin{equation}
\Gamma_{12}(\tau) = \frac{1}{T} \int_0^T E_1(t+\tau)E_2^*(t) dt
\end{equation}

Note that if $E_1(t+\tau)$ and $E_2(t)$ are not correlated, the integral will tend to zero as $T$ increases (but not, if they are correlated).

Defining the intensity fluctuations $\Delta I$ as:

\[
\Delta I_1(t) = I_1(t) - \langle I_1 \rangle
\qquad \Delta I_2(t) = I_2(t) - \langle I_2 \rangle ,
\]

one obtains:

\begin{equation}
  \langle \Delta I_1(t) \Delta I_2(t) \rangle = \langle I_1 \rangle \langle I_2 \rangle |\gamma_{12}|^2,
\label{intcorr4}
\end{equation}

since $\langle \Delta I \rangle = 0$.  These equations hold for linearly polarized light; for unpolarized light, a factor $1/2$ enters on the right-hand side.

Here, $|\gamma_{12}|$ equals the classical visibility as measured from the minimum and maximum intensities $I_{max}$ and $I_{min}$ in an amplitude interferometer:

\begin{equation}
V =  |\gamma_{12}| = \frac{I_{max}-I_{min}}{I_{max}+I_{min}} ,
\end{equation} 

which thus ranges from 0 (destructive interference fringes) to 1 (when $I_{min}=0$). 

This illustrates the sensitivity of amplitude interferometers to atmospheric or optical imperfections: an effective drop of fringe contrast and visibility from maximum to zero may be caused by a phase change of just $\lambda$/2, requiring an instrumental stability to within a small fraction of one wavelength.  An intensity interferometer measures $|\gamma_{12}|^2$ with a certain electronic time resolution.  This quantity remains positive irrespective of atmospheric or optical disturbances although -- since realistic time resolutions do not reach down to optical coherence times -- it may get strongly diluted relative to the full value it would have had in the case of a hypothetical `perfect' temporal resolution (shorter than the light-wave period).  For realistic values of nanoseconds, this dilution typically amounts to several orders of magnitude and thus the directly measurable quantity $|\gamma_{12}|^2$ becomes quite small.  This is the reason why very good photon statistics are required, implying large flux collectors.{\footnote{For a given electronic time resolution, this dilution is smaller for lower-frequency electromagnetic radiation (with longer coherence time), and for long-wavelength infrared and radio, this additional variability due to fluctuations in the signal itself (i.e., not caused by any detector imperfections) is more easily measured, and there known as 'wave noise'.}}

\section{Optical aperture synthesis}

The original intensity interferometer at Narrabri used two telescopes placed at different separations  $r$ to deduce angular sizes of stars from the observed value of $|\gamma_{12}(r)|^2$, analogous to what can be measured with a two-element amplitude interferometer.  Systems with multiple telescopes and different baselines enable more complete image reconstructions, be it either amplitude or intensity interferometers.

Interferometric image reconstruction and aperture synthesis was pioneered in radio.  For details, see, e.g., Taylor et al.\ (1999) or Thompson et al.\ (2001); applications to the optical are treated by Glindemann (2011), Labeyrie et al.\ (2006), Millour (2008) or Saha (2011); here we recall the basics:

The separation vector between a pair of telescopes in a plane perpendicular to the line of observation, the $(u,v)-$plane, is $\mathbf{r_1-r_2}$, so that for the optical wavelength $\lambda$, $\mathbf{r_1-r_2} = (u \lambda, v \lambda)$.  If the telescopes are not in such a plane, also a third coordinate enters: the time-delay $w$ for the propagation of light along the line of sight to the source; $\mathbf{r_1-r_2} = (u \lambda, v \lambda, w)$.

With the angular coordinate positions of the target $(l, m)$, one can deduce the following expression for the correlation function $\Gamma_{12} = \langle E(\mathbf{r}_1) E^*(\mathbf{r}_2) \rangle$:

\begin{equation}
\Gamma(u,v) = \iint I(l, m) e^{-2 \pi i (ul+vm)} dl dm.
\end{equation}

This equation represents the van Cittert-Zernike theorem, which states that the quantity measured by an [amplitude] interferometer for a given baseline is a component of the Fourier transform of the surface intensity distribution of the source.  This Fourier transform can be inverted:

\begin{equation}
I_{\nu}(l,m) = \iint V(u, v) e^{2 \pi i (ul+vm)} du dv,
\label{vancittert}
\end{equation}

where $V(u,v)$ equals the normalized value of $\gamma(u,v)$.  Thus, by using multiple separations and orientations of interferometric pairs of telescopes, one can sample the $(u,v)-$plane and reconstruct the source image with a resolution equal to that of a telescope with a diameter of the longest baseline.  This, of course, is the technique of aperture synthesis.

In intensity interferometry, however, a complication enters in that the correlation function for the electric field, $\gamma_{12}$, is not directly measured, but only the square of its modulus, $|\gamma_{12}|^2$.  Since this does not preserve phase information, the {\it{direct}} inversion of the above equation is not possible.

With only two telescopes, the original intensity interferometer could only carry out a quite sparse sampling of the $(u,v)-$plane.  While this was sufficient to determine the angular extent of stellar disks, and to search for the flattened shapes of rapid rotators, more elaborate image analyses were not practical.

This limitation will be removed in intensity interferometry carried out with large arrays of Cherenkov telescopes.  With some 50 or more flux collectors, the possible number of baselines becomes enormous; $N$ telescopes can form $N(N-1)/2$  baselines, reaching numbers in the thousands (even if possibly periodic telescope locations might make several of them redundant).  Since such telescopes are fixed on the ground, they trace out ellipses in the $(u,v)-$plane, as a source moves across the sky.  With proper signal handling, all successive measures of $|\gamma_{12}|^2$ can be allocated to their specific $(u,v)-$coordinates, producing a highly filled $(u,v)-$plane, with a superior coverage of projected orientations across the source image.  As will be discussed below, such complete data coverage indeed enables reconstruction of the phases of the Fourier components, and thus permits full image reconstructions.

For large numbers of telescopes, another advantage of intensity interferometry becomes obvious.  Since telescopes are connected only electronically, there is (in principle) no loss of signal when synthesizing any number of baselines between any pairs of telescopes: the digital signal from each telescope is simply copied electronically.  By contrast, amplitude interferometry in the optical (as opposed to radio) requires optical beams of starlight between telescopes since the very high optical frequency (combined with rapid phase fluctuations in chaotic light) precludes its amplification with retained phase information.{\footnote{Although quantum-optical procedures can be envisioned to realize even this (Gottesman et al.\ 2011).}}  In order to obtain the many baselines needed for efficient aperture synthesis (such as realized in radio), starlight from each telescope would need to be split and sent to several other telescopes, each combination with its own delay line system.  While such ambitious arrangements can be made for a moderate number of telescopes (e.g., Creech-Eakman et al.\ 2010), the complexity (and the dilution of light between different baselines) rapidly increases if any greater number of telescopes would be engaged.

\section{Cherenkov telescope arrays}

The largest complex currently planned is the CTA (2012), envisioning on the order of 50-100 telescopes with various apertures between about 5 and 25 meters, distributed over an edge-to-edge distance of some 2 km.  Baselines in currently existing Cherenkov arrays do not exceed some 200 meters, and their achievable angular resolution largely overlaps with that feasible with existing amplitude interferometers (although one could observe in the blue or violet, where the contrast of many stellar features is expected to be higher).  Some experiments in connecting pairs of Cherenkov telescopes for intensity interferometry have already been carried out (see below), and although some observations could be made already with existing facilities, any significant leap in optical astronomy will require larger arrays. 

\begin{figure}
\centering
\includegraphics[width=11cm, angle=90]{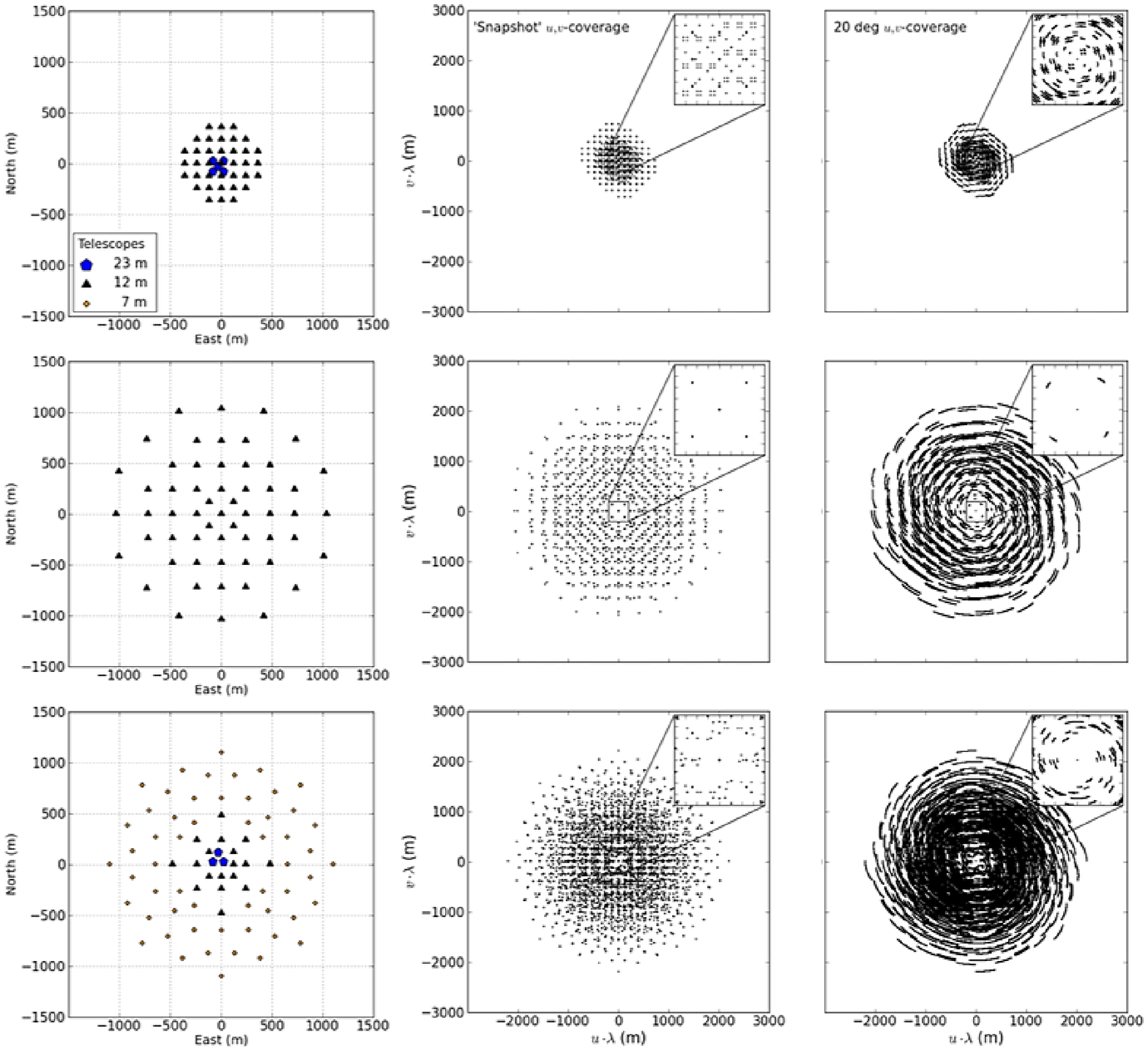}
\caption{Left: Telescope placements on the ground for different configurations studied for the planned Cherenkov Telescope Array (details in Table \ref{configurationstable}).  Middle column: $(u,v)-$plane coverages at an instant in time, for a star in the zenith. Upper right-hand squares expand the central 400{$\times$}400 m area.  Right: $(u,v)-$plane coverages for a star moving from the zenith through 20 degrees to the west.  The numerous telescopes enable a huge number of baseline pairs which largely fill the entire $(u,v)-$plane.}
\label{fig3}
\end{figure}

A number of candidate array layouts for the CTA were considered within its design study (Actis et al.\ 2011; Bernl{\"o}hr et al.\ 2008; CTA 2012; Hermann 2010) of which examples representing different types of layouts are in Figure \ref{fig3} and Table \ref{configurationstable}.  For interferometry, large telescope separations, i.e., long baselines, measure high-frequency components, corresponding to small structures on the target, while short baselines sample the low frequencies.  For an Earth-bound interferometer (in a plane perpendicular to the line of observation) with a baseline $\mathbf{B} = (B_{\mathrm{North}}, B_{\mathrm{East}})$ the associated coordinates in the Fourier $(u,v)-$plane are $(u,v) = \frac{1}{\lambda}(B_{\mathrm{North}}, B_{\mathrm{East}})$.

For stationary telescopes, the projected baselines, $\mathbf{B}_p$, will change while the target of observation moves across the sky, with each telescope pair tracing out an ellipse in the Fourier plane according to the following expression (S{\'e}gransan 2007):

\begin{scriptsize}
\begin{equation}
\label{rotationsynth}
\begin{pmatrix}
u \\
v \\
w
\end{pmatrix}
=
\frac{1}{\lambda} \mathbf{B}_p =
\frac{1}{\lambda} \begin{pmatrix}
-\sin l \sin h & \cos h & \cos l \sin h\\
\sin l \cos h \sin \delta + \cos l \cos \delta & \sin h \sin \delta & -\cos l \cos h \sin \delta + \sin l \cos \delta \\
-\sin l \cos h \cos \delta + \cos l \sin \delta & -\sin h \cos \delta & \cos l \cos h \cos \delta + \sin l \sin \delta
\end{pmatrix}
\begin{pmatrix}
B_{north} \\
B_{east}\\
B_{up}
\end{pmatrix}
\end{equation}
\end{scriptsize}

where $l$ is the latitude of the telescope array, and $\delta$ and $h$ are the declination and hour angles of the star. The $w$ component corresponds to the time delay in the wavefront arrival time between the two telescopes (dependent on also the elevation difference of the telescopes, $B_{up}$).  The extensive coverage of the $(u,v)-$plane that results from the Earth's rotation enables the synthesis in software of a very large telescope and -- of course -- is the very principle used in much of radio interferometry.

Figure \ref{fig3} illustrates these capabilities for arrays of Cherenkov telescopes.  Here, three among the potential layouts considered for CTA are taken as examples for qualitatively somewhat different telescope arrangements.  One is a compact configuration; another a sparse and rather uniform one; and a third with telescopes of different sizes grouped with successively different spacings.  The large telescopes (23\,m diameter in this concept) near center offer the best sensitivity for lower-energy gamma rays, the medium-size (12\,m) ones cover a larger area, while the small ones (7\,m) are spread widely to better record the Cherenkov light pool induced by the highest-energy gammas.

The latter type of layouts seem to lie close to those currently favored for the CTA layout and, as seen in Figure \ref{fig3}, already short observations of just an hour or so may cover much of the $(u,v)-$plane.

\begin{figure}
\centering
\includegraphics[width=7.5cm, angle=90]{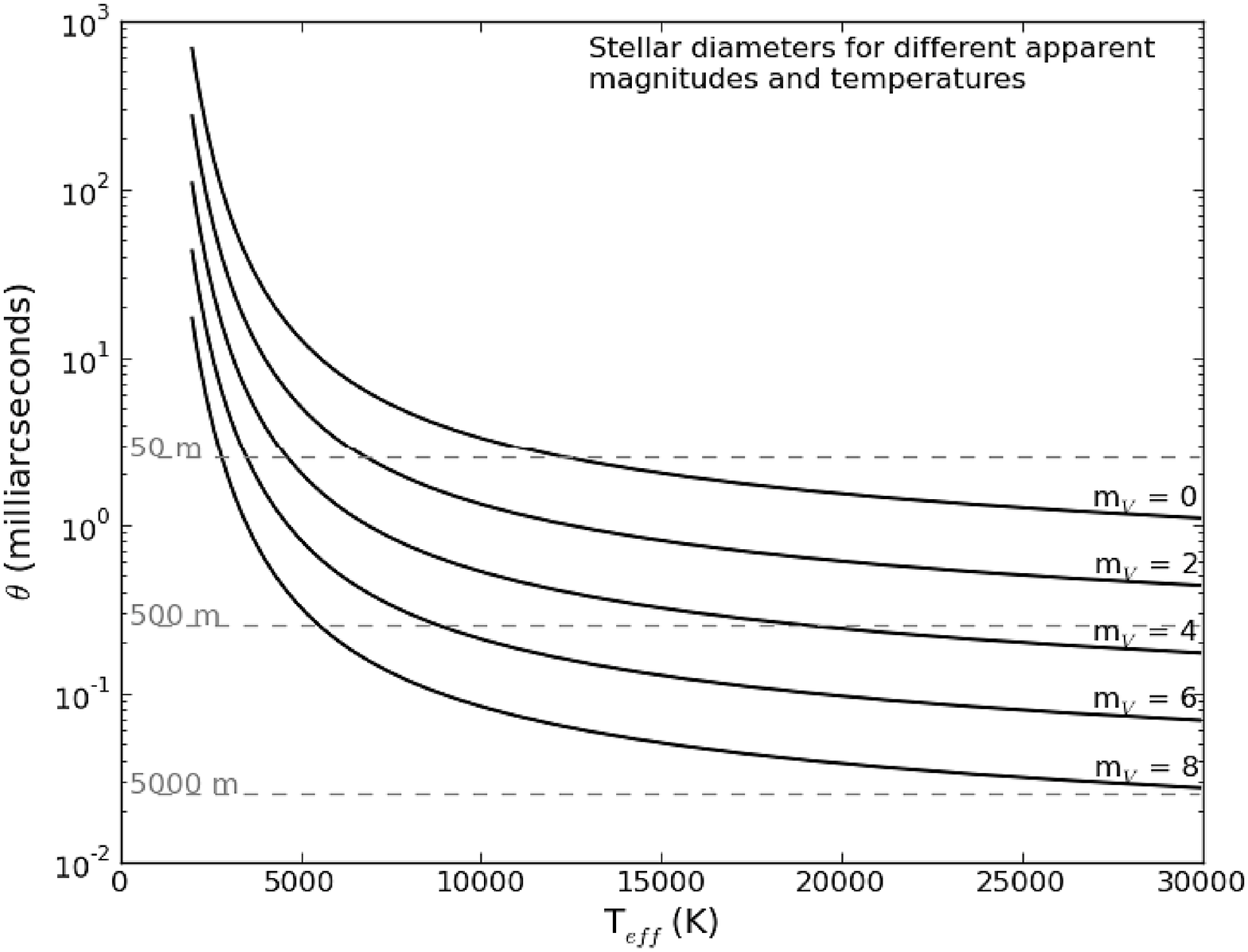}
\caption{Relationship between stellar diameter and effective temperature for different apparent magnitudes. Stars are assumed to be blackbodies with uniform circular disks, observed in the V band (centered on $\lambda$~545\,nm).  Dashed lines show baselines at which different diameters are resolved, i.e., where the first minimum of the spatial coherence function is reached.  \label{fig4}}
\end{figure}

\section{Signal-to-noise in intensity interferometry} \label{5}

No other current instrument in astronomy measures the second-order coherence of light and, since its noise properties differ from those of other observations, they are essential to understand in defining realistic observing programs.  For one pair of telescopes, the signal-to-noise ratio (Hanbury Brown 1974; Twiss 1969) is given in a first approximation by:

\begin{equation}
(S/N)_{RMS} = A \cdot \alpha \cdot n \cdot |\gamma_{12}(\mathbf{r})|^2 \cdot \Delta f^{1/2} \cdot (T/2)^{1/2} 
\end{equation}

where $A$ is the geometric mean of the areas (not diameters) of the two telescopes; $\alpha$ is the quantum efficiency of the optics plus detector system; $n$ is the flux of the source in photons per unit optical bandwidth, per unit area, and per unit time; $|\gamma_{12}(\mathbf{r})|^2$ is the second-order coherence of the source for the baseline vector $\mathbf{r}$, with $\gamma_{12}(\mathbf{r})$ being the mutual degree of coherence.  $\Delta f$ is the electronic bandwidth of the detector plus signal-handling system, and $T$ is the integration time.

Most of these parameters depend on the instrumentation,  but $n$ depends on the source itself, being a function of also its radiation temperature.  For a given number of photons detected per unit area and unit time, the signal-to-noise ratio is better for sources where those photons are squeezed into a narrower optical band.  The method is based upon two-photon correlations and more photons inside one optical coherence volume imply a higher probability for detecting two of them simultaneously.  Alternatively, from a classical wave-optics point of view, a narrower passband implies a more monochromatic source with a longer coherence time, and smaller loss of temporal coherence during the electronic integration time.

This property implies that (for a flat-spectrum source) the S/N is {\it{independent of}} the width of the optical passband, whether measuring only the limited light inside a narrow spectral feature or a much greater broad-band flux.  Although at first perhaps somewhat counter-intuitive, the explanation is that realistic electronic resolutions of nanoseconds are very much slower than the temporal coherence time of broad-band light (perhaps 10$^{-14}$~s).  While narrowing the spectral passband does decrease the photon flux, it also increases the temporal coherence by the same factor, canceling the effects of increased photon noise.  This property was exploited already in the Narrabri interferometer by Hanbury Brown et al.\ (1970) to identify the extended emission-line volume from the stellar wind around the Wolf-Rayet star $\gamma^2$~Vel.  The same effect could also be exploited for increasing the signal-to-noise by observing the same source simultaneously in multiple spectral channels, a concept foreseen for the once proposed successor to the original Narrabri interferometer (Davis 1975; Hanbury Brown 1979; 1991).  

To be a feasible target for kilometric-scale interferometry, any source must provide both a significant photon flux, and be small enough for its structures to produce significant visibility over such long baselines (Figure \ref{fig4}).  This implies that the method is particularly sensitive to hotter sources: cool ones would have to be large in extent to give a sizeable flux, but then will be spatially resolved already over short baselines.  Seen alternatively, for stars with the same angular diameter but decreasing temperature (thus decreasing fluxes), telescope diameter must be increased in order to maintain the same S/N.  When the star is resolved by a single aperture, the S/N begins to drop (the spatial coherence of the light decreases), and no gain results from larger mirrors.

Given that the electronic signal bandwidth cannot realistically be much higher than about a gigahertz, the temporal coherence of the light is diluted (compared to a hypothetical `full' time resolution of maybe 10$^{-14}$~s), and a significant photon flux is required in order to measure the second-order coherence to a good precision.  Calculations, simulations, and extrapolations from work with the past Narrabri instrument demonstrate that, for realistic Cherenkov telescope performance, under normal night-sky conditions, the limiting visual magnitude for determining the angular size of a continuum source will be on order m$_V$=\,9 (LeBohec \& Holder 2006).  The magnitude limits are discussed further below, and of course any such number is only approximate as it might be pushed by employing larger flux collectors with better optics (taking in less sky backlground), more senitive detectors, higher signal bandwidth, and/or simultaneously observing in multiple spectral channels.

Since the signal-to-noise ratio does not depend on the width of the spectral passband, it follows that a source with bright emission lines may be observed in just those lines to enhance the S/N to a level corresponding to the emission-line radiation temperature, while the integrated light from the source could be fainter than those magnitude limits.  Already in work preceding the Narrabri intensity interferometer, estimates of possible S/N (assuming then foreseen detector sensitivity and electronic bandwidth; measuring in one optical passband; using circular telescope mirrors, integrating for 1 hour) were given by Hanbury Brown \& Twiss (1958; their Figure 6; see also Twiss 1969) as function of stellar temperature: about 200 for 10,000\,K, reaching 1000 for 20,000\,K.  Such numbers will of course improve with better instrumentation but, for any given electronic performance, stars cooler than a certain temperature will not give any sensible signal-to-noise ratio, no matter how bright the star, or how large the telescopes.  Aspects of achievable S/N are further discussed by Foellmi (2009) and Schulz \& Gupta (1998).

In principle, the signal could be enhanced by increasing the electronic bandwidth (up to that of the light itself, of 10$^{15}$\,Hz or so), but then one would essentially have re-created an amplitude interferometer with all its requirements to control optical and electronic delays to within 10$^{-15}$\,s or less, equivalent to the light-travel distance over a fraction of an optical wavelength, exactly the requirement that intensity interferometry was set out to circumvent in the first place.

\section{Simulated observations in intensity interferometry}

To obtain quantitative measures of what can be observed using realistic detectors on actual or planned air Cherenkov telescopes, a series of simulations were carried out.

\subsection{Numerical simulations} 

An intensity interferometer using two photon-counting detectors $A$ and $B$ and a digital correlator measures the squared modulus of the complex degree of coherence of the light:

\begin{equation}
\label{intcorr}
|\gamma|^2 = \frac{\langle \Delta I_1 \Delta I_2 \rangle}{\langle I_1 \rangle \langle I_2 \rangle}
\end{equation}

or, in a discrete form:

\begin{equation}
g^{(2)} = \frac{N_{AB}}{N_A N_B} N,
\label{disccorr}
\end{equation}

where $N_A$ and $N_B$ are the number of photons detected in $A$ and $B$ respectively, $N_{AB}$ is the number of joint detections (i.e., the number of time intervals in which both detectors record a photon), and $N$ is the number of sampled time intervals.  Since a strict Monte-Carlo simulation would be computationally very demanding, a simplified procedure was used by generating random numbers $N_A$, $N_B$ and $N_{AB}$, and inserting these into Eq.\ \eqref{disccorr}. These will be Poisson distributed random variables\footnote{In practice, the measurement time is always long enough for the Poisson distributions to be adequately approximated as normal distributions.} with mean values $\mu_A = P_A\cdot N$, $\mu_B = P_B\cdot N$ and $\mu_{AB} = P_{AB}\cdot N$. Here, $P_A$ and $P_B$ are the probabilities of detecting a photon in $A$ and $B$ respectively, within a small time interval $\Delta t$, and $P_{AB}$ is the probability of a joint detection within $\Delta t$.

These probabilites can be written out in terms of variables depending only on the instrumentation and the target of study:
\begin{align}
P_A &= \alpha_A \langle I_A \rangle \Delta t \\
P_B &= \alpha_B \langle I_B \rangle \Delta t \\
P_{AB} &= P_A P_B + \alpha_A \alpha_B \langle I_A \rangle \langle I_B \rangle |\gamma_{AB}|^2 {\tau_c}{\Delta t} 
\end{align}

\begin{figure}
\centering
\centering
\includegraphics[width=6cm,angle=90]{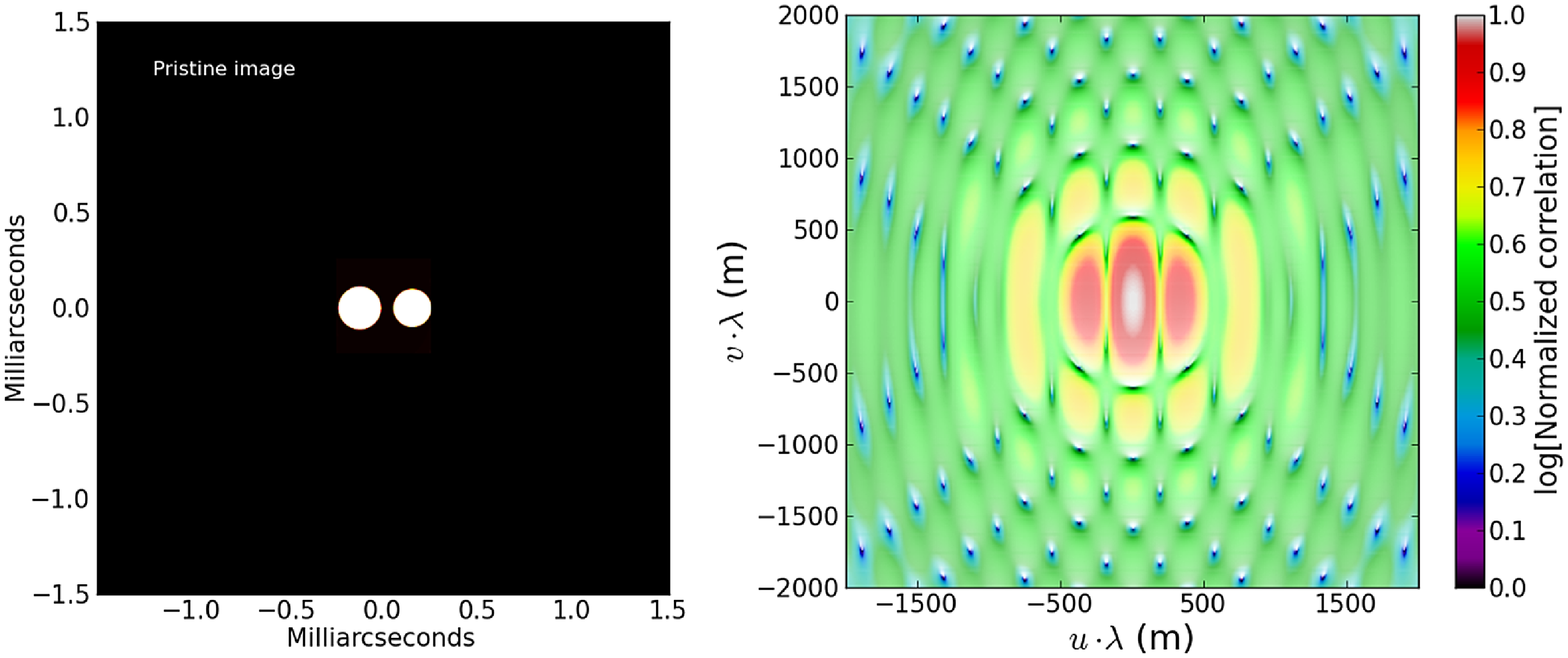}
\caption{Image of a close binary star with 200 and 150\,$\mu$as diameter components, used to simulate observations, and the (logarithmized) magnitude of its Fourier transform.  This noise-free pattern is what would be measured by a perfect interferometer of projected size 2000{$\times$}2000\,m.  Corresponding patterns in later figures cover only some part of this  $(u,v)-$plane (due to finite extent of the telescope array on the ground), and become noisy for fainter sources and finite integration times. }
\label{fig:binarysim}
\end{figure}

Here $\alpha$ denotes the quantum efficiency of the detectors, $\langle I \rangle$ is the mean light intensity, $\tau_c$ is the coherence time of the light (determined by the wavelength and optical passband) and $\gamma_{AB}$ is the degree of optical spatial coherence (proportional to the Fourier transform of the target image, assuming telescope sizes to be small compared to the spatial structure in this transform).   Such simulations were carried out for the various telescope-array configurations and for various assumed sources.  Here, examples are shown for a close binary star with components taken as uniform disks of diameters 200 and 150\,$\mu$as.  Both the pristine image and the pristine Fourier transform in the $(u,v)-$plane are shown in Figure \ref{fig:binarysim}.  Across the Fourier plane, the magnitude of various patterns varies greatly.  To enhance the visibility of also fainter structures (and to better see the effects of noise), the Fourier-plane figures use a logarithmic scaling and a shading to enhance the contrast  (the exact numerical values of the measured correlations are not significant in this context).

Also results from the simulated observations are mostly given as such Fourier-plane images rather than full image reconstructions.  The simulated observations produced values at many different discrete locations in the $(u,v)-$plane, which were used in a linear interpolation to obtain the Fourier magnitude over a regular grid.  This image format makes the effects of noise and changing telescope arrangements easier to interpret since it is independent of the performance of algorithms for image reconstruction or data analysis.  As discussed below, optimal image reconstruction is a developing research topic of its own: even though reconstructed images do reflect the capability of the simulated telescope array, some reconstructions are still limited by the algorithms used.  By contrast, the information recovered in the $(u,v)-$plane is independent of such algorithm performance.

\subsection{Array layouts and limiting magnitudes}

Perhaps a first question is how faint are the sources that can be observed?  Figure \ref{magnitudes} shows the results of varying the brightness of the target.  Here, simulations  have the star starting in the zenith and moving 50 degrees to the west in steps of 5 degrees, with an integration time of 2 hours for each time step (thus a total of 22 hours, spanning several nights).  The assumed observation wavelength is 500\,nm and the filter bandwidth 1\,nm{\footnote{This small value was chosen due to computational limits in handling high photon-count rates; however -- as noted above -- the S/N is in principle independent of the optical bandwidth.}.  Results from the somewhat optimistic values taken for the quantum efficiency (70\,\%), and the time resolution of the digital detectors (1\,ns discretization step) could be scaled down for using fewer telescopes, lesser efficiency, or coarser time resolution (or scaled up, for multiple spectral channels).  Since the angular sizes of the stars are kept constant, their effective temperatures depend on the assumed visual brightness as per Figure \ref{fig4}.  For example, a star of m$_V$=\,5 would have an effective temperature around 10,000\,K. 

\begin{figure}
\centering
\includegraphics[width=12cm, angle=90]{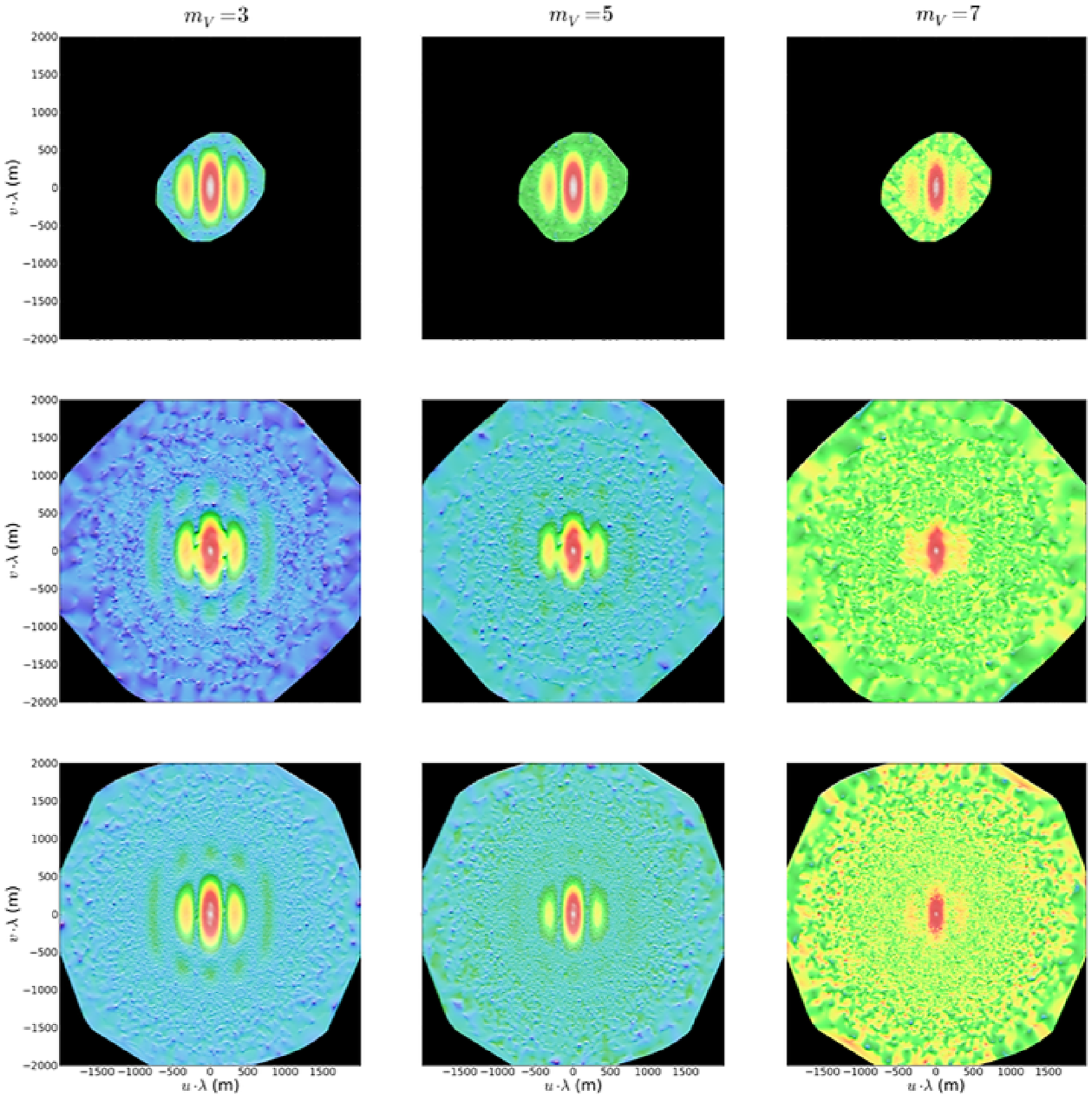}
\caption{Simulated observations of a binary star, for assumed visual magnitudes m$_V$=\,3, 5, and 7.  The resulting magnitudes of the Fourier transforms in the $(u,v)-$plane are shown, for (top to bottom) the three telescope configurations in Figure \ref{fig3} and Table \ref{configurationstable}.   A comparison to the pristine Fourier transform in Figure \ref{fig:binarysim} shows how much of the information on the source geometry that is retained.}
\label{magnitudes}
\end{figure}

A clear difference among the three telescope configurations is caused by the different extent of the arrays, and thus the differently long baselines that can be synthesized.   Configuration 1 (top) samples the central parts of the Fourier plane very densely and provides a field of view that is large enough to image the overall shape of any brighter star (of typical diameter 1--2\,mas). However, since longer baselines are lacking, it is unable to resolve details smaller than $\sim$\,200 $\mu$as.  Configuration 2 (center), on one hand, provides long baselines out to 2180\,m, permitting studies of detailed structures, down to $\sim\,$50$\mu$as at 500\,nm but, on the other hand, its shortest baseline is 170\,m, which means that any structures larger than $\sim$\,0.75\,mas will be lost, making it unsuitable for most stellar sources.  Configuration 3 (bottom) seems to provide the best of two worlds. It has baselines short enough to measure the shapes of the disks of most stellar objects, while still providing very long baselines and a very good resolution (cf.\ Table~{\ref{configurationstable}}).  Effects of different array geometries are discussed further by Jensen et al.\ (2010).

Examining the noise level and geometry of the `observed' Fourier transform in the $(u,v)-$plane, it seems that while the m$_V$=\,5 source still produces some  information on also its finer structures (seen as higher-order diffraction rings), such details start to disappear for m$_V$=\,7.  From such simulations, we conclude that a realistic limiting magnitude for {\it{two-dimensional}} imaging with such a large array is around m$_V$=\,6, in general agreement with previous estimates (e.g., LeBohec \& Holder 2006).  If only some {\it{one-dimensional}} measure is sought (e.g., a stellar diameter or limb darkening), the data can be averaged over all position angles, and the limiting magnitude will become correspondingly fainter.

\subsubsection{Comparing to the original Narrabri interferometer}

At present, no astronomical intensity interferometer is operating, constraining the practical verification of such simulations.  However, we can use the same procedure to simulate past observations with the classical intensity interferometer at Narrabri for Sirius.  Figure \ref{sirius} shows both original measurements from Narrabri, and simulated observations of a uniform stellar disk with properties corresponding to those of Sirius (diameter $\theta=5.6$\,mas, $T_{eff}=10,000$\,K, m$_V=-$1.46).  The simulation was made for parameters analogous to those at Narrabri: two 6.5\,m reflectors; $\lambda$\,450\,nm; detector quantum efficiency 10\,\%; time resolution 10\,ns, observing 40 hours for each data point.

\begin{figure}
\centering
\includegraphics[width=5.5cm,angle=90]{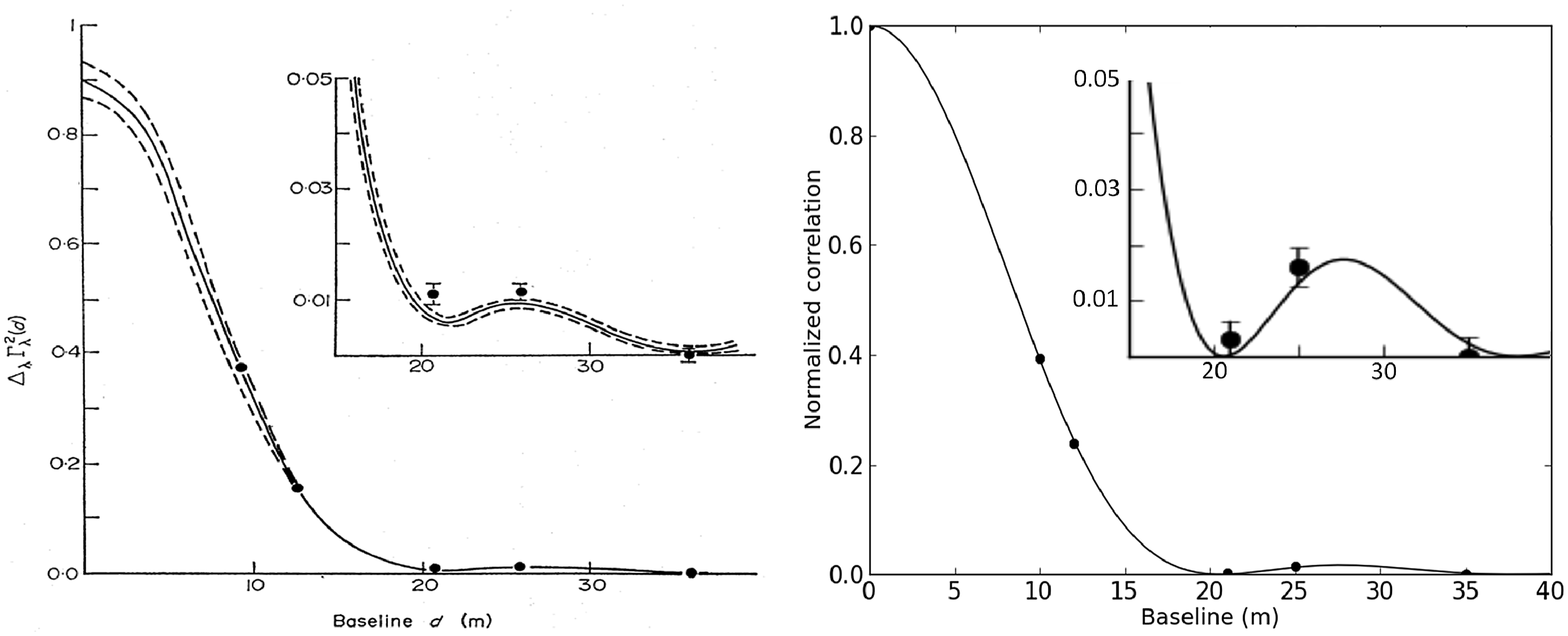}
\caption{Left: Observations of Sirius made with the Narrabri intensity interferometer by Hanbury Brown et al.\ (1974b), compared to current simulations with similar input parameters. The lines show the theoretical correlation curves; omission of stellar limb darkening causes a larger amplitude of the simulated secondary maximum at right.  Error bars in the simulation plot were calculated using Eq.\ \eqref{signal_to_noise}.}
\label{sirius}
\end{figure}

The results look very similar (a small difference in the correlation curves is due to the limb darkening incorporated in the stellar model used to fit the Narrabri observations).  Our signal-to-noise estimates were obtained through a Monte-Carlo method: A large number $n$ of correlation measurements were made for zero baseline (full spatial coherence) and stored in a list $\mathbf{C}_1$.  A second list $\mathbf{C}_2$ was produced for a very long baseline (zero coherence).  The vector $\mathbf{C} = \mathbf{C}_1-\mathbf{C}_2$ is then regarded as the signal, and the signal-to-noise ratio taken as the mean of the signal divided by the standard deviation of the signal:

\begin{equation}
\label{signal_to_noise}
\frac{S}{N} = \frac{\bar{\mathbf{C}}} { \sqrt{\frac{1}{n} \sum_{i=1}^n (\mathbf{C}_i-\bar{\mathbf{C}})^2} }.
\end{equation}

The considerable similarity between the classical observations and our simulations appears to confirm their credibility.  Still, we note that there are limits in how exact comparisons can be made with past measurements.  For a bright star such as Sirius, the digital detectors in our current simulations start to saturate (i.e., they approach one photon per time-resolution interval) and some sort of light-attenuation device has to be used, unless the spectral filter is very narrow.  Thus, the filter bandwidth affects the signal-to-noise for bright stars in a way that is different from the past use of analog detectors at Narrabri.

\section{Imaging with intensity interferometry}
\label{reconstruction}

As already mentioned, an intensity interferometer directly measures only the absolute magnitudes of the respective Fourier transform components of the source image that cover the $(u,v)-$plane, while the phases are not directly obtained.  Such Fourier magnitudes can well be used by themselves to fit model parameters such as stellar diameters, stellar limb darkening, binary separations, circumstellar disk thicknesses, etc., but actual images cannot be directly computed from the van~Cittert-Zernicke theorem, Eq.\eqref{vancittert}.  A two-component interferometer, such as the classical one at Narrabri, provides only very limited coverage of the $(u,v)-$plane, and it seems doubtful whether much more can be extracted.  However, a multi-component interferometer offers numerous baselines, and gives an extensive coverage of the $(u,v)-$plane (cf. Figure {\ref{fig3}}).  In this case, the $(u,v)-$plane may be largely filled (Figures {\ref{fig:binarysim}} and {\ref{magnitudes}}), and it is already intuitively clear that the information contained there must place rather stringent constraints on the source image.

\subsection{Phase reconstruction}
\label{phase_reconstruction}

A number of techniques have been developed for recovering the phase of a complex function when only its magnitude is known.  One method specifically intended for intensity interferometry was worked out by Holmes \& Belen'kii (2004) and Holmes et al.\ (2010) for one and two dimensions, respectively.  Once a sufficient coverage of the Fourier plane is available, and phase recovery has been performed, a study on imaging capabilities can be carried out.  Such studies by Nu{\~n}ez et al.\ (2010; 2012ab) applied Cauchy-Riemann based phase recovery to reconstruct images from simulated intensity interferometry observations.  Also rather  complex images can be reconstructed, demonstrating that imaging at the submilliarcsecond scale is feasible.  A limitation that remains is the non-uniqueness between the image and its mirrored reflection.  Some earlier discussions on the potential information content are by Bates (1969) and Kurashov \& Khoroshkov (1976).

It is simpler to first understand phase retrieval in one dimension where one approximates the continuous Fourier transform $F(x)$ by a discrete one, expressing it as a polynomial in the complex variable $z$, where $z\equiv e^{imk_0\Delta x \Delta \theta}$.  The theory of analytic functions can then be applied to this polynomial, in particular using the Cauchy-Riemann equations in polar form to relate the phase and the log-magnitude along the real or imaginary axes\footnote{The Cauchy-Riemann equations can be applied because $F(z)$ is a polynomial in $z$, where $z\equiv e^{i\phi}$. These relate the phase $\Phi$ and the log-magnitude $\ln R$ along the real or imaginary axes.}.  One can show that the phase differences along the radial direction in the complex plane directly relate to the differences in the logarithm of the magnitude; see Holmes \& Belen'kii (2004) for details, so that integrating the Cauchy-Riemann equations directly does not immediately solve for the phase: phase differences along the purely real or imaginary axes are not directly available from the data.

Since $z$, the independent variable of the Fourier transform, has modulus equal to 1, data for the phase differences that we seek are only available along the unit circle in the complex plane ($|z|=1$).  The procedure to find the phase consists in first assuming a plausible solution form, then taking differences in the radial direction of the complex plane, and finally fitting the data to the radial differences of the assumed solution.  A general form of the phase can be postulated by noting that it is a solution of the Laplace equation in the complex plane (applying the Laplacian operator on the phase and using the Cauchy-Riemann equations yields zero). Since the phase differences are known along the radial direction in the complex plane, we can take radial differences of the general solution and then fit the log-magnitude differences (available from the data) to these.

One can think of this one-dimensional reconstruction as the phase estimation along a single slice through the origin in the Fourier plane.  An extension to two dimensions can be made by combining multiple 1-D reconstructions, with many directions of the slices in the $(u,v)-$plane, while ensuring mutual consistency between successive 1-D slices.  This approach benefits from the high S/N ratio that is present near the origin of the Fourier plane (assuming a suitable telescope layout), and can also benefit from additional information that may be available at low frequencies near the origin of the Fourier plane.  Such information may comprise continuity or a separate measurement using a conventional telescope, for example.  These 1-D phase reconstructions are inherently ambiguous to within an overall flip and hence one must resolve any flips of these in order to ensure that a reasonable image is formed. This is done using continuity by comparing the correlation of neighboring radial phase reconstructions and, if necessary, flipping to maximize such correlation, in sequence in the Fourier plane of the image; phasing is ensured by overlap of 1-D reconstructions near the origin (Holmes et al.\ 2010).

There are other possible approaches for phase retrieval, such as Gerchberg-Saxton phase retrieval, Generalized Expectation Maximization, and other variants of the Cauchy-Riemann approach (Holmes et al.\ 2010), or even simple genetic algorithms.  The problem of optimal image reconstruction under various noise levels indeed recurs in various imaging applications (unrelated to astronomy) and there is a significant literature on this and related issues (Fienup 1982; Hurt 1989; Schulz \& Snyder 1992; Schulz \& Voelz 2005).  Since it is a research topic on its own (like, perhaps, image reconstruction was in the early days of radio aperture synthesis), the examples shown here serve only to demonstrate a proof of concept of realizing two-dimensional image reconstruction from intensity interferometry and not the appearance of optimally reconstructed images.

\begin{figure}
\centering
\includegraphics[width=6.5cm,angle=90]{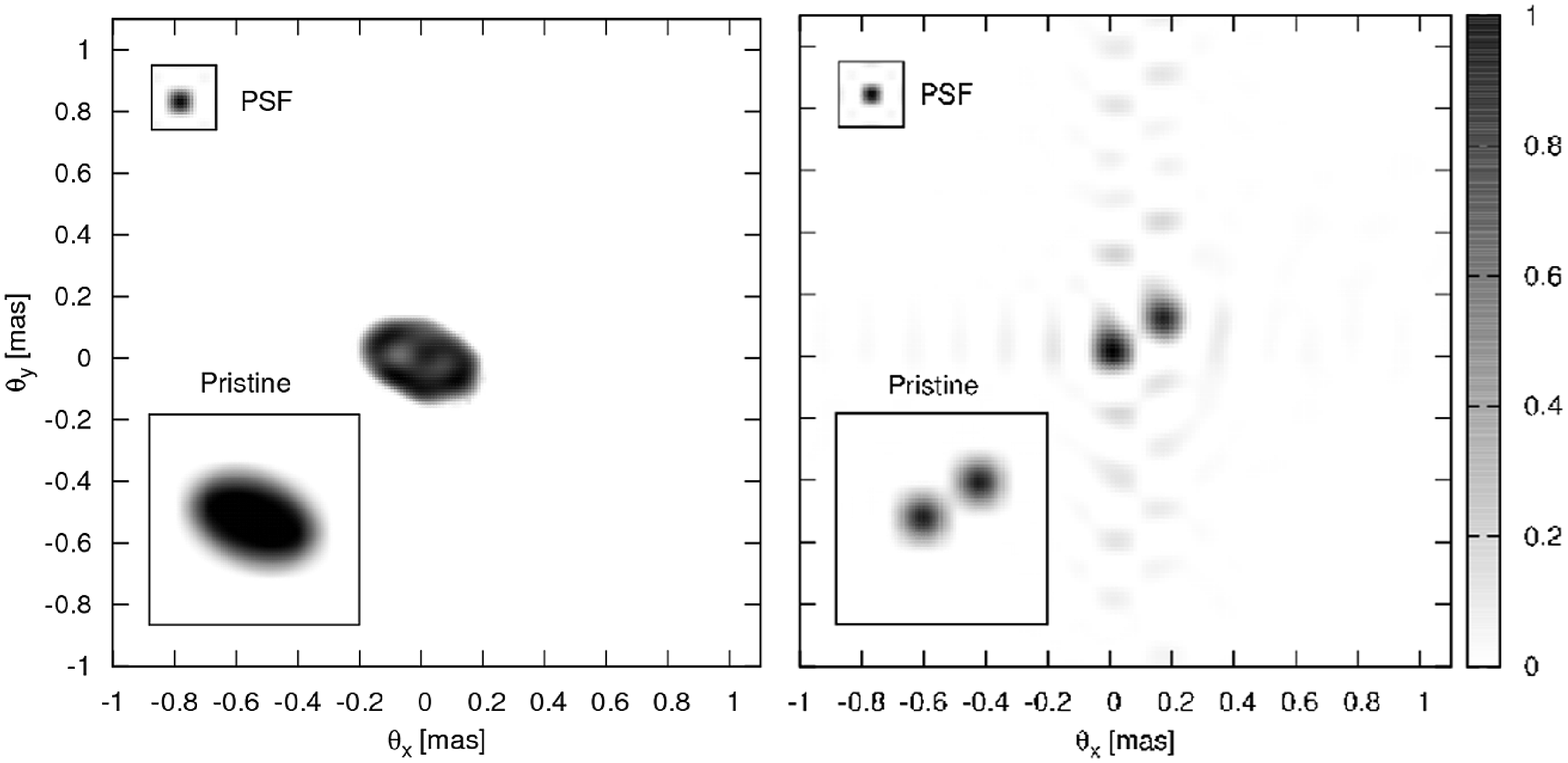}
\caption{Left: Simulated and reconstructed images of an oblate rapidly rotating star of m$_V$=\,3 after 10 hours of observation time with a large Cherenkov telescope array. Right: Images of a close binary of magnitude m$_V$=\,6, for 50 hours of observation.  The `pristine' images show the sources convolved with the theoretical point-spread-function of the array, showing the theoretical resolution limit of some 60\,{$\mu$as.  The reconstructions were obtained without any preliminary assumptions of what the images would look like. }}
\label{oblate_and_binary}
\end{figure}

\subsection{Image reconstruction}
\label{images}

This Cauchy-Riemann approach has been tested (Nu{\~n}ez et al.\ 2010; 2012ab) on data from simulated observations of objects such as oblate rotating stars, binary stars, and stars with brighter or darker regions, and here a few representative examples are shown.   The left part of Figure {\ref{oblate_and_binary}} shows a reconstruction of an oblate rotating star of magnitude m$_V$=\,3, after 10 hours of observation on a 97-telescope array (each telescope with light-collecting area 100 m$^2$, baselines between 50 and 1500\,m; an early CTA concept).  The semi-major and semi-minor axes of the pristine ellipse are 200\,$\mu$as and 120~$\mu$as respectively.  The right-hand part of the same Figure {\ref{oblate_and_binary}} illustrates the reconstruction of an m$_V$=\,6 binary star after 50 hours of exposure time.  In this Figure, the `pristine' images are the original ones, convolved with the theoretical point spread function of the array, thus showing what could in principle be achieved with a `perfect' image reconstruction.

These reconstructed images have undergone post-processing, analogous to such applied in conventional amplitude interferometry.  The image is then iteratively modified by small increments to maximize the agreement between the data (squared modulus of Fourier magnitude) and the magnitude of the Fourier transform of the reconstructed image.  We use the MiRA (`Multi-aperture image reconstruction algorithm') software developed by Thi{\'e}baut (2009). Such analysis depends strongly on the starting image, which can be provided by the phase recovery presented above.  However, it should be noted that for applications such as estimating stellar oblateness or modeling the orbit of a binary star, a more robust result is obtained by fitting to the original data in the $(u,v)-$plane, without involving any image reconstruction.  

Such `final' images are preceded by the retrieval of `raw' images.  Since there are different means of fitting the data for use in the reconstruction algorithm, also the character of the possible artifacts or noise in the resulting reconstructions may vary.  For example, if reconstructing the phase by taking horizontal one-dimensional slices of the Fourier magnitude (and relating them to each other with a single vertical slice) noise patterns might preferentially appear in a vertical direction.  As for the structure (bumps) within the oblate star in Figure \ref{oblate_and_binary}, these start to appear when either the star becomes bright enough, or enough exposure time is supplied so that information other than the first lobe in the Fourier magnitude becomes significant.  At high spatial frequencies, the signal is still noisy, e.g., the signal-to-noise ratio at baselines greater than 600\,m approaches unity.  Noisy high-frequency portions in the Fourier plane may cause fictitious structure to appear, most likely since most of that high-frequency information is used to reconstruct a dark background (several milliarcseconds in extent), with a central bright region.  Such examples illustrate that further improvements must be possible for reconstructing high frequencies (i.e., structures within stars).  For further discussions, see Nu{\~n}ez et al.\ (2010; 2012ab) and Holmes et al.\ (2010).   

\clearpage

\begin{table}
\begin{center}
\caption{Candidate sources from The Bright Star Catalogue (Hoffleit \& Warren 1995): 35 stars brighter than m$_V$=\,2 {\it{or}} hotter than T$_{eff}$\,= 25,000\,K. Those whose angular diameters were measured already with the Narrabri intensity interferometer (Hanbury Brown et al.\ 1974a) are marked with an asterisk (*).  Angular diameters $\theta$ are in mas, stellar rotational velocities V$_{rot}$ in km\,s$^{-1}$, and effective temperatures T$_{eff}$ in K.  
\label{source_table}}

\tiny{
\begin{tabular}{lrcclccl}
Name & HR & $\theta$  & V$_{rot}$  & Spectr.\  & T$_{eff}$  &  m$_V$ & Notes \\
Achernar, $\alpha$~Eri & 472 & 1.9 & 250 & B3 Ve & 15 000 & 0.46 & High V$_{rot}$, *  \\
Rigel, $\beta$~Ori & 1713 & 2.4 & 30 & B8 Iab &  9 800 & 0.12 & Emission-line star, * \\
$\lambda$~Lep & 1756 &  & 70 & B0.5 IV &	28 000 &	4.29 & \\
Bellatrix, $\gamma$~Ori & 1790 & 0.7 & 60 & B2 III &	21 300 &	1.64 & Variable, * \\
Elnath, $\beta$~Tau \\=~$\gamma$~Aur [sic] & 1791 &	1.5 & 70 & B7 III & 13 500 & 1.65 & Binary system \\
$\upsilon$~Ori & 1855 &	& 20 & B0 V & 28 000 & 4.62 & Variable \\
HD 36960 & 1887 & & 40 & B0.5 V & 26 000 & 4.78 & Binary system \\
Alnilam, $\epsilon$~Ori & 1903 & 0.7 & 90 & B0 Iab & 18 000 & 1.7 & Emission-line star, * \\
$\mu$~Col & 1996 & & 150 & O9.5 V & 33 000 & 5.17 & \\
$\beta$~CMa & 2294 & 0.5 & 35 &	B1 II-III & 23 000 & 1.98 &	$\beta$~Cep-type variable, * \\
Alhena, $\gamma$~Gem & 2421 &	1.4 & 30 & A0 IV & 9 100 & 1.93 & * \\
S~Mon & 2456 & & 60 & O7 Ve & 26 000 & 4.66 & Pre-main-sequence \\
Sirius, $\alpha$~CMa & 2491 & 5.9 & 10 & A1 V &  9 100 & --1.46 & * \\
EZ CMa & 2583 & & & WN4 & 33 000 & 6.91 & Highly variable W-R star \\
Adara, $\epsilon$~CMa & 2618 & 0.8 & 40 & B2 Iab & 20 000 & 1.5 & Binary, * \\
Naos, $\zeta$~Pup & 3165 & 0.4 & 210 & O5 Ia & 28 000 & 2.25 & BY Dra variable, * \\
$\gamma^2$~Vel & 3207 & 0.4 & & WC8 & 50 000 & 1.78 & Wolf-Rayet binary, * \\
& & & &  O7.5 & 35 000 & & \\
$\beta$~Car & 3685 & 1.5 & 130 &	A2 IV & 9 100 & 1.68 & * \\
Regulus, $\alpha$~Leo & 3982 & 1.4 & 330 &	B7 V & 12 000 & 1.35 & High V$_{rot}$, * \\
$\eta$~Car & 4210 & 5.0 & & peculiar & 36 000 & 6.21 & Extreme object, variable \\
Acrux, $\alpha^1$~Cru & 4730 & & 120 & B0.5 IV & 24 000 & 1.33 & Close binary to  $\alpha^2$~Cru \\
Acrux, $\alpha^2$~Cru & 4731 & &	200 & B1 V & 28 000 & 1.73 & Close binary to $\alpha^1$~Cru \\
$\beta$~Cru & 4853 & 0.7 & 40 & B0.5 IV &	23 000 &	1.25 & $\beta$~Cep-type variable, * \\
$\epsilon$~UMa & 4905 &	& 40 & A0 p & 9 500 & 1.77 &  $\alpha^2$~CVn-type variable \\
Spica, $\alpha$~Vir & 5056 & 0.9 & 160 & B1 III-IV & 23 000 & 0.98 & $\beta$~Cep-type variable \\
Alcaid, $\eta$~UMa & 5191 & $<$~2 & 200 & B3 V & 18 000 & 1.86 & Variable \\
$\beta$~Cen & 5267 & 0.9 & 140 &	B1 III & 23 000 & 0.61 & $\beta$~Cep-type variable \\
$\tau$~Sco & 6165 & & 25 & B0.2 V & 26 000 & 2.82 & \\
$\lambda$~Sco & 6527 &  & 160 & B2 IV+ & 21 000 & 1.63 & $\beta$~Cep-type variable \\
Kaus Australis, $\epsilon$~Sgr & 6879 & 1.4 &	140 & B9.5 III & 9 800 & 1.85 & Binary, * \\
Vega, $\alpha$~Lyr & 7001 & 3.2 & 15 & A0 V & 9 100 & 0.03 &	* \\
Peacock, $\alpha$~Pav & 7790 & 0.8 & 40 &	B2 IV & 19 000 & 1.94 & Spectroscopic binary, * \\
Deneb, $\alpha$~Cyg & 7924 & 2.2 & 20 & A2 Iae & 9 300 & 1.25 & Variable \\
$\alpha$~Gru & 8425 & 1.0 & 230 & B6 V & 13 000 & 1.74 & * \\
Fomalhaut, $\alpha$~PsA & 8728 & 2 & 100 & A4 V & 9 300 & 1.16 & With imaged exoplanet, * \\
\end{tabular}
}
\end{center}
\end{table}

\section{The new stellar physics}

With optical imaging approaching resolutions of tens of microarcseconds (and with also a certain spectral resolution), we are moving into novel and previously unexplored parameter domains, enabling new frontiers in astrophysics.  However, pushing into these domains requires attention not only to optimizing the instrumentation but also to a careful choice of targets to be selected.  These must be both astronomically interesting and realistic to observe with currently planned facilities.  With a foreseen brightness limit of perhaps m$_V$=\,6 or 7, and with sources of a sufficiently high brightness temperature, initial observing programs have to focus on bright stars or stellar-like objects (Dravins et al.\ 2010).

\subsection{Hot and bright sources} 

Among the about 9,000 objects in the Bright Star Catalogue (Hoffleit \& Warren 1995), some 2,600 objects are both hotter than 9,000\,K {\it{and}} brighter than m$_V$=\,7, among which the brightest and hottest should be those easiest to observe.  Table~\ref{source_table} lists such a subset of 35 stars brighter than m$_V$=\,2 {\it{or}} hotter than T$_{eff}$~= 25,000 K (most effective temperatures were approximated from measured $B-V$ colors, using a polynomial fit to values from Bessell et al.\ 1998). 

Naturally, this list of potential targets partially overlaps with those that were selected for diameter measurements already with the Narrabri interferometer; in Table \ref{source_table} those are marked with asterisks.   Of course, with longer baselines, stars can be not only spatially resolved but one may start analyzing structures on and around them; compared to past Narrabri targets, this list is biased more towards hotter stars, with typically smaller diameters, as appropriate for longer baselines.

To quantify the total number of sources for which sensible intensity correlations can be measured with a large array of air Cherenkov telescopes, Nu{\~n}ez et al.\ (2012a) extracted data from the Jean-Marie Mariotti Center stellar diameters catalog (Lafrasse et al.\ 2010), finding that $\sim$\,1000 stars should be detectable within 1\,h, $\sim$\,2500 stars in 10\,h and $\sim$\,4300 stars within 50 hours of observation.  Even if the exact numbers depend on assumed instrumental efficiencies, this shows that thousands of sources will be accessible to currently planned telescope complexes.

\subsection{Primary targets} 

\subsubsection{Rapidly rotating stars} 

Rapidly rotating stars are normally hot and young ones, of spectral types O, B, and A; some are indeed rotating so fast that the effective gravity in their equatorial regions becomes very small (at critical rotation even approaching zero), and easily enables mass loss or the formation of circumstellar disks.  Rapid rotation causes the star itself to become oblate, and induces gravity darkening.  A theorem by von Zeipel (1924) states that the radiative flux in a uniformly rotating star is proportional to the local effective gravity and implies that equatorial regions are dimmer, and polar ones brighter.  Spectral-line broadening reveals quite a number of early-type stars as rapid rotators and their surface distortion was looked for already with the Narrabri interferometer, but not identified due to then insufficient signal-to-noise levels (Hanbury Brown et al.\ 1967b; Johnston \& Wareing 1970).

A number of these have now been studied with amplitude interferometers (van Belle 2012).  By measuring diameters at different position angles, the rotationally flattened shapes of the stellar disks are determined.  For some stars, also their asymmetric brightness distribution across the surface is seen, confirming the expected gravitational darkening and yielding the inclination of the rotational axes.  Aperture synthesis has permitted the reconstruction of images using baselines up to some 300\,m, corresponding to resolutions of 0.5\,mas in the near-infrared H-band around  $\lambda$\,1.7\,$\mu$m (Zhao et al.\ 2009).

Two stars illustrate different extremes: Achernar ($\alpha$~Eridani) is a highly deformed Be-star (V$_{rot}$sin\,i = 250 km\,s$^{-1}$; $>$\,80 \% of critical).  Its disk is the flattest so far observed -- the major/minor axis ratio being 1.56 (2.53 and 1.62\,mas, respectively); and this projected ratio is only a lower value -- the actual one could be even more extreme (Domiciano de Souza et al.\ 2003).  Further, the rapid rotation of Achernar results in an outer envelope seemingly produced by a stellar wind emanating from the poles (Kervella \& Domiciano de Souza 2006; Kervella et al.\ 2009).  There is also a circumstellar disk with H{$\alpha$}-emission, possibly structured around a polar jet (Kanaan et al.\ 2008). The presence of bright emission lines is especially interesting: since the S/N of an intensity interferometer is independent of the spectral passband, studies in the continuum may be combined with observations centered at an emission line.

Going to the other extreme, Vega ($\alpha$~Lyrae, A0~V) has been one of the most fundamental stars for calibration purposes but its nature has turned out to be quite complex.  First, space observations revealed an excess flux in the far infrared, an apparent signature of circumstellar dust.  Later, optical amplitude interferometry showed an enormous (18-fold) drop in intensity at $\lambda$\,500\,nm from stellar disk center to the limb, indicating that Vega is actually a very rapidly rotating star which just happens to be observed nearly pole-on.  The true equatorial rotational velocity is estimated to 270\,km\,s$^{-1}$; while the projected one is only 22\,km\,s$^{-1}$ (Aufdenberg et al.\ 2006; Peterson et al.\ 2006).  The effective polar temperature is around 10,000\,K, the equatorial only 8,000\,K.  The difference in predicted ultraviolet flux between such a star seen equator-on, and pole-on, amounts to a factor five, obviously not a satisfactory state for a star that should have been a fundamental standard.

Predicted classes of not yet observed stars are those that are rotating both rapidly and differentially, i.e., with different angular velocities at different depths or latitudes.  Such stars could take on weird shapes, midway between a donut and a sphere (MacGregor et al.\ 2007).  There exists quite a number of hot rapid rotators with diameters of one mas or less, and clearly the angular resolution required to reveal such stellar shapes would be 0.1\,mas or better, requiring kilometric-scale interferometry for observations around $\lambda$\,400 nm.

\subsubsection{Circumstellar disks} 

Rapid rotation lowers the effective gravity near the stellar equator which enables centrifugally driven mass loss and the development of circumstellar structures.  Be-stars make up a class of rapid rotators with dense equatorial gas disks; the `e' in `Be' denotes the presence of emission in H$\alpha$ and other lines.  Observations indicate the coexistence of a dense equatorial disk with a variable stellar wind at higher latitudes, and the disks may evolve, develop and disappear over timescales of months or years (Porter \& Rivinius 2003). 

The detailed mechanisms for producing such disks are not well understood, although the material in these decretion (mass-losing) disks seems to have been ejected from the star rather than accreted from an external medium.  The rapid rotation of the central B star certainly plays a role (Townsend et al.\ 2004).   Some Be-stars show outbursts, where the triggering mechanism is perhaps coupled to non-radial pulsations.  Some of their disks have been measured with amplitude interferometers, e.g.,  $\zeta$~Tau (Carciofi et al.\ 2009; Gies et al.\ 2007).  A related group is the B[e] one, where emission is observed in forbidden atomic lines from [Fe II] and other species.  A few of those stars are within realistic magnitude limits (e.g., HD\,62623 = $l$~Pup of m$_V$=\,4.0).

\subsubsection{Winds from hot stars} 
     
The hottest and most massive stars (O-, B-, and Wolf-Rayet types) have strong and fast stellar winds that are radiatively driven by the strong photospheric flux being absorbed or scattered in spectral lines formed in the denser wind regions.  Not surprisingly, their complex time variability is not well understood.  Stellar winds can create co-rotating structures in the circumstellar flow in a way quite similar to what is observed in the solar wind. These structures have been suggested as responsible for discrete absorption components observed in ultraviolet P Cygni-type line spectra. 

Rapid stellar rotation causes higher temperatures near the stellar poles, and thus a greater radiative force is available there for locally accelerating the wind.  In such a case, the result may be a poleward deflection of wind streamlines, resulting in enhanced density and mass flux over the poles and a depletion around the equator (opposite to what one would perhaps `naively' expect in a rapidly rotating star).  Surface inhomogeneities such as cooler or hotter starspots cause the local radiation force over those to differ, leading to locally faster or cooler stellar-wind streamers which may ultimately collide, forming co-rotating interaction regions.  Further, effects of magnetic fields are likely to enter and -- again analogous to the case of the solar wind -- such may well channel the wind flow in complex ways (ud-Doula \& Owocki 2002).

\subsubsection{Wolf-Rayet stars and their environments} 

Being the closest and brightest Wolf-Rayet star, and residing in a binary jointly with a hot O-type star,  $\gamma^2$~Velorum is an outstanding object for studies of circumstellar interactions.  The dense Wolf-Rayet wind collides with the less dense but faster O-star wind, generating shocked collision zones, wind-blown cavities and eclipses of spectral lines emitted from a probably clumpy wind (Millour et al.\ 2007; North et al.\ 2007). The bright emission lines enable studies in different passbands, and already with the Narrabri interferometer, Hanbury Brown et al.\ (1970) could measure how the circumstellar emission region (seen in the C\,III-IV feature around $\lambda$~465\,nm) was much more extended than the continuum flux from the stellar photosphere, and seemed to fill much of the Roche lobe between the two components of the binary. 

A few other binary Wolf-Rayet stars with colliding winds are bright enough to be realistic targets.  One is WR~140 (m$_V$=\,6.9, with bright emission lines), where the hydrodynamic bow shock has been followed with milliarcsecond resolution in the radio, using the Very Long Baseline Array (VLBA), revealing how the bow-shaped shock front rotates as the orbit progresses during its 7.9\,yr period (Dougherty et al.\ 2005).

\subsubsection{Blue supergiants and related stars} 

Luminous blue variables occupy positions in the Hertzsprung-Russell diagram adjacent to those of Wolf-Rayet stars, and some of these objects are bright enough to be candidate targets, e.g., P~Cyg (m$_V$=\,4.8).  Luminous blue variables possess powerful stellar winds and are often believed to be the progenitors of nitrogen-rich WR-stars.  Rigel ($\beta$~Ori; B8\,Iab) is the closest blue supergiant (240\,pc). It is a very dynamic object with variable absorption/emission lines and oscillations on many different timescales.  Actually, the properties of Rigel resemble those of the progenitor to supernova SN1987A.

$\beta$~Centauri (B1\,III) is a visual double star, whose primary component is a spectroscopic binary with two very hot, very massive, pulsating and variable stars in a highly eccentric orbit ($e$\,=\,0.82; Ausseloos et al.\ 2002; Davis et al.\ 2005).  Its binary nature was first revealed with the Narrabri interferometer (Hanbury Brown et al.\ 1974a), then measuring a significantly lower intensity correlation than expected from a single star.  The formation history of such massive and highly eccentric systems is not understood; a few others are known but $\beta$~Cen is by far the brightest one (also the brightest variable of the $\beta$~Cep type), and thus a prime target.

A most remarkable luminous blue variable is $\eta$~Carinae, the most luminous star known in the Galaxy.  It is an extremely unstable and complex object which has undergone giant eruptions with huge mass ejections during past centuries.  The mechanisms behind these eruptions are not understood but, like Rigel, $\eta$~Car may well be on the verge of exploding as a core-collapse supernova.  Interferometric studies reveal asymmetries in the stellar winds with enhanced mass loss along the rotation axis, i.e., from the poles rather than from the equatorial regions, resulting from the enhanced temperature at the poles that develops in rapidly rotating stars (van Boekel et al.\ 2003; Weigelt et al.\ 2007). 

\subsubsection{Interacting binaries} 

Numerous stars in close binaries undergo interactions involving mass flow, mass transfer and emission of highly energetic radiation: indeed many of the bright and variable X-ray sources in the sky belong to that category.  However, to be a realistic target for intensity interferometry, they must also be optically bright, which typically means B-star systems.  

One well-studied interacting and eclipsing binary is $\beta$~Lyrae (Sheliak; m$_V$=\,3.5).  The system is observed close to edge-on and consists of a B7-type, Roche-lobe filling and mass-losing primary, and an early B-type mass-gaining secondary. This secondary appears to be embedded in a thick accretion disk with a bipolar jet seen in emission lines, causing a light-scattering halo above its poles.  The donor star was initially more massive than the secondary, but has now shrunk to about 3\,M$_\odot$, while the accreting star has reached some 13\,M$_\odot$.  The continuing mass transfer causes the 13-day period to increase by about 20 seconds each year (Harmanec 2002). 

Using the CHARA interferometer with baselines up to 330\,m, the $\beta$~Lyr system has been resolved in the near-infrared H and K bands (Zhao et al.\ 2008).  The images resolve both the donor star and the thick disk surrounding the mass gainer, 0.9\,mas away.  The donor star appears elongated, thus demonstrating the photospheric tidal distortion due to Roche-lobe filling.  Numerous other close binaries invite studies of mutual irradiation, tidal distortion, limb darkening, rotational distortion, gravity darkening, and oscillations.  These include Spica ($\alpha$~Vir; m$_V$=\,1.0; primary B1\,III-IV); the pre-main-sequence 15~Mon (S~Mon; m$_V$=\,4.7; O7\,V(f) + O9.5\,Vn); HD\,193322; m$_V$=\,5.8 (primary O9\,V); { $\delta$}~Sco (m$_V$=\,2.3; primary B0~IVe); {$\delta$}~Ori (m$_V$=\,2.2; O9\,II + B0\,III); and the complex of stars in the Trapezium cluster, e.g., $\theta^1$~Ori\,C (m$_V$=\,5.1; primary O6pe), and others.

Another class of interacting stars is represented by Algol ($\beta$~Persei; m$_V$=\,2.1), a well-known eclipsing binary in a triple system, where the large and bright primary $\beta$~Per\,A (B8\,V) is regularly eclipsed by the dimmer K-type subgiant $\beta$~Per~B, for several hours every few days.  It could appear as a paradox that the more massive  $\beta$~Per\,A is still on the main sequence, while the presumably coeval but less massive  $\beta$~Per\,B already has evolved into a subgiant: significant mass transfer must have occurred from the more massive companion and influenced stellar evolution.  Algol is also a flaring radio and X-ray source, and analyses of its variability suggest that to be related to magnetic activity which apparently affects the mass transfer and the accretion structure.  Possibly, not only the cooler (solar-type) star is magnetically active, but magnetic fields are generated also by hydrodynamically driven dynamos inside the accretion structures (circumstellar disks or annuli).  The disk and stellar fields interact, with magnetic reconnection causing energy release in flares and acceleration of relativistic particles (Retter et al.\ 2005).  As mentioned already for Be-type stars, magnetic fields can in addition channel the gas flows in the system and generate quite complex geometries.

\subsubsection{Novae and eruptive variables} 

Transient sources may reach brightnesses that are adequate for interferometric observation.  In particular, about a dozen novae are detected in the Galaxy each year and every few years some may reach naked-eye brightness (e.g., Nova Cygni 1975 reached m$_V$=\,1.7 and V1280 Scorpii had m$_V$=\,3.8 in 2007).  These cataclysmic explosions caused by the thermonuclear runaway fusion of hydrogen, following its accretion onto a white-dwarf surface, display a wide variety of complex and incompletely understood phenomena likely to show significant spatial structure.  

Imaging a bright nova with intensity interferometry using a large telescope array could be especially attractive since a near-complete $(u,v)-$plane coverage would be assured already after a short observation, enabling a monitoring of the evolving shape of the expanding fireball.  To be practically observable, however, requires the source to be not only visually bright but also sufficiently hot.  These conditions should be satisfied if catching a nova still in its early fireball state, when the ejected hot gases are in the process of initial cooling.  In those early phases, the material is very hot, 30,000\,K or more, comparable to that of the hottest ordinary stars (e.g., Munari et al.\ 2008).  During the subsequent expansion of the nova photosphere towards maximum total brightness, the overall temperature drops to 10,000\,K and below, at some stage possibly becoming marginal for interferometry.  After maximum light, however, the opacity of the expanding shell drops and one may start seeing into deeper and hotter layers, with temperatures again in a range of perhaps 30,000-60,000\,K.  Further, the blast-wave ejecta may locally be at a very high temperature (also likely regions of X-ray and gamma-ray emission; Nelson et al.\ 2012; Orlando \& Drake 2012).  Novae display rich emission-line spectra whose different conditions of formation reflect different regions and depths of the ejected envelope (e.g., Shore et al.\ 2011; 2012).  Given that the signal-to-noise ratio in intensity interferometry is independent of the spectral bandpass, one can envision simultaneous monitoring of the nova eruption in multiple spectral lines to deduce its three-dimensional structure.

Also some classes of other eruptive variables might be candidates.  Visually bright supernovae are very rare events, but -- as evidenced by SN1987A -- if they do occur in the nearby Universe, their brightness can be appreciable.

\begin{figure}
\centering
\includegraphics[width=4.5cm, angle=90]{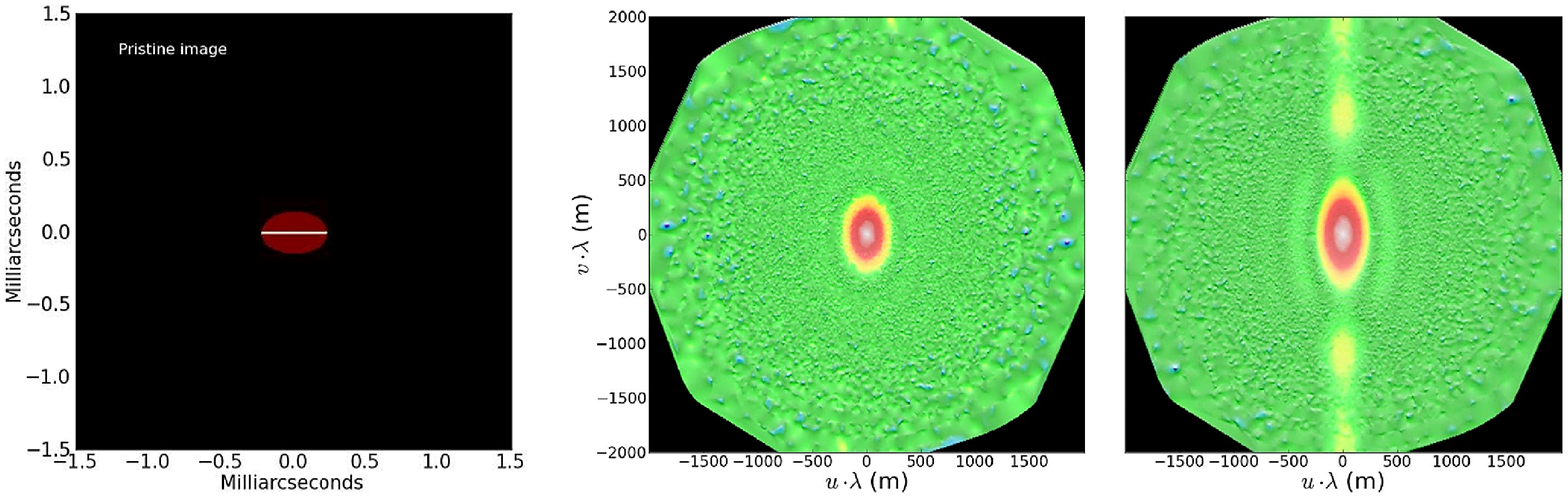}
\caption{Simulated observations of a rotationally flattened star with a very thin (10\,$\mu$as across) emission-line disk seen edge-on.  Left: Assumed pristine image.  Center: Simulated observations of the magnitude of the two-dimensional Fourier transform of the source's intensity distribution in continuum light, as sampled by a large number of telescopes.  The flattening of the stellar disk is visible as an asymmetry in the $(u,v)-$plane.  Right: The same, but for a narrow-bandpass filter centered on the He\,I emission line, showing the distinct signature of a narrow equatorial disk.  \label{flat_star_disk}}
\end{figure}

\subsection{Observing programs} 

The most promising targets for early intensity interferometry observations thus appear to be relatively bright and hot, single or binary O-, B-, and WR-type stars with their various circumstellar emission-line structures.  The expected diameters of their stellar disks are typically on the order of 0.2--0.5\,mas and thus lie (somewhat) beyond what can be resolved with existing amplitude interferometers.  However, several of their outer envelopes or disks extend over a few mas and have been resolved with existing facilities, thus confirming their existence and providing hints on what types of features to expect when next pushing the resolution by another order of magnitude.  Also, when observing at short wavelengths (and comparing to amplitude interferometer data in the infrared), one will normally observe to a different optical depth in the source, thus beginning to reveal also its three-dimensional structure.

Also some classes of somewhat cooler objects are realistic targets.  Some rapidly rotating A-type stars of temperatures around 10,000\,K should be observable for their photospheric shapes (maybe one could even observe how the projected shapes change with time, as the star moves in its binary orbit, or if the star precesses around its axis?).  Stars in the instability strip of the Hertzsprung-Russell diagram, of spectral types around F and temperatures below 7,000\,K, undergo various types of pulsations.  For example, the classic Cepheid $l$~Car (m$_V$=\,3.4) was monitored at  $\lambda$\,700 nm with the SUSI interferometer over a 40\,m baseline, finding its mean diameter of 3.0 mas to cyclically vary over its 35-day pulsation period with an amplitude of almost 20 \% (Davis et al.\ 2009). 

However, the diameters of such brighter Cepheids (typically 1-3\,mas) can be resolved already at modest baselines, and those that would require kilometric baselines are too faint for presently foreseen intensity interferometry.  Nevertheless, several such stars are expected to undergo non-radial pulsations, with sections of the stellar surface undulating in higher-order modes.  The modulation amplitudes in temperature and white light presumably are modest (not likely to realistically be detectable) but the corresponding velocity fluctuations could perhaps be observed.  If the telescope optics permit an adequate collimation of light to enable measurements through a very narrow-band spectral filter centered on a stronger absorption line of 50\,\% residual intensity, say, the local stellar surface will appear at that particular residual intensity (if at rest relative to the observer), but will reach full continuum intensity if the local velocities have Doppler-shifted the absorption line outside the filter passband.  If such spatially resolved observations of stellar non-radial oscillations can be realized, they would provide highly significant input to models of stellar atmospheres and interiors (Cunha et al.\ 2007; Jankov et al.\ 2001; Schmider et al.\ 2005).

\subsection{More complex sources}

In the previous discussion, simulated observations were shown for rather simple sources but also various more complex geometries have been numerically simulated, assuming configurations envisioned for currently planned Cherenkov telescope facilities.  An example is shown in Figure \ref{flat_star_disk} for a rapidly rotating and rotationally flattened star, (m$_V$=\,6; T$_{eff}$=\,7,000\,K), some 0.4\,mas across, seen equator-on, with a very thin (10\,$\mu$as) disk visible in the He\,I emission line at $\lambda$\,587 nm, assumed to be six times stronger than the local continuum (a plausible value for Be or B[e] emission-line stars; e.g., Lamers et al.\ 1998).  For an electronic time resolution of 1 ns and a detector quantum efficiency of 70\,\%, data were assumed to be integrated for 10 hours with a telescope configuration analogous to one being discussed for the CTA, similar to previous calculations.  The center and right-hand panels illustrate the roles of different baselines: The flattened stellar disk is resolved already by the innermost few-hundred-meter baselines while the signal (`diffraction pattern') from the very narrow (10\,$\mu$as) emission disk clearly continues even beyond the assumed longest baselines, and it is obvious that significant information on its geometry can be extracted.   Some full image reconstructions were also been carried out but are not shown here because (similarly to what was discussed above) they still are more limited by the performance of particular reconstruction algorithms rather than by intrinsic interferometric capabilities.  However, for samples of reconstructed images, see Nu{\~n}ez et al.\ (2010; 2012ab).

\section{Observing in practice} 

In this Section, we examine various practical issues in carrying out actual observations in intensity interferometry, concerning aspects of the telescopes, detectors, data handling and the scheduling of observations.

\subsection{Optical $e$-interferometry} 

Electronic combination of signals from multiple telescopes is currently being done for long-baseline radio interferometry, where remote radio antennas are electronically connected to a common signal-processing station via optical fiber links in so-called $e$-VLBI.  This is feasible due to the relatively low radio frequencies (MHz-GHz); a corresponding optical phase-resolved signal (THz-PHz) could not be managed but the much slower intensity-fluctuation signal (again MHz-GHz) is realistic to transmit, thus enabling an electronic connection of also optical telescopes.
    
Several authors have noted this potential fo optical $e$-interferometry, and a number of suggestions exist in the recent literature.  Dravins et al.\ (2005) and Dravins (2008) point at the potential of electronically combining multiple subapertures of extremely large telescopes, especially for observations at short optical wavelengths.  Ofir \& Ribak (2006abc) evaluate concepts for multidetector intensity interferometers, and even space-based intensity interferometry has been proposed (Hyland et al.\ 2007; Klein et al.\ 2007), exploiting the possibility to combine signals off-line from each component telescope, thus relaxing the requirement for spacecraft orientation and orbital stability.  With a reference star within the field of view, intensity interferometry might even be used for astrometry, possibly in searches for exoplanets (Hyland 2005; 2007).

\subsection{Performance of Cherenkov telescopes} 

As already mentioned, the specifications of air Cherenkov telescopes are remarkably similar to the requirements for intensity interferometry.  The signals to be measured for intensity interferometry have much in common to those of atmospheric Cherenkov flashes: nanosecond time structure and relatively short optical wavelengths.  Most probably, the same types of very fast photon-counting detectors can be used, although the sources to be observed are much brighter, and the data handling has to allow for continuous integrations (rather than trigger-based acquisition of short data bursts). 

\subsubsection{Image quality} 

Even if the technique as such does not require good optical quality, and permits also rather coarse flux collectors with point-spread functions of several arcminutes, issues arise from unsharp stellar images: in particular an increased contamination by the background light from the night sky.  Although this background light does not contribute any net intensity-correlation signal, it increases the photon-counting noise, especially when observing under moonlight conditions.

While any reasonable optical quality should be adequate for intensity interferometry as such, the magnitude m$_V$ of the faintest stars that can be studied may depend on the optical point spread function.  Two extreme sky brightness situations can be: (a) dark observatory sky with $\sim$\,21.5 m$_V$/arcsec$^2$; (b) sky with full Moon, $\sim$\,18 m$_V$/arcsec$^2$.  The contamination expected from the sky background then results in a flux equal to stellar magnitude m$_V$ $\sim$\,9.4 (a) and 5.9 (b) for a 5 arcmin diameter field, and m$_V$ $\sim$\,12.9 (a) and 9.4 (b) for 1 arcmin diameter.

A larger point-spread function also takes in other sky events (meteors, distant flashes of lightning, etc.), and may preclude the use of small-sized semiconductor detectors of possibly higher quantum efficiency.

\subsubsection{Isochronous optics} 

For Cherenkov light observations, a large field of view is desired.  In most optical systems, the image quality deteriorates away from the optical axis, and to mitigate this, various optical solutions are used.  Many current telescopes have the layout introduced by Davies \& Cotton (1957), whose primary reflector forms a spherical structure, giving smaller aberrations off the optical axis compared to a parabolic design.  The primary mirror is made up of numerous reflector facets, all of the same focal length $f$, arranged on a sphere of radius $f$.   

This has the consequence that the telescope optics become not isochronous, i.e., photons originally on the same wavefront, but striking different parts of the entrance aperture may not arrive to the focus at exactly the same time.  As noted in Section \ref{5} above, the signal-to-noise ratio improves with electronic bandwidth, i.e., the time resolution with which stellar intensity fluctuations can be measured.  The time spread induced by a non-isochronous telescope acts like an `instrumental profile' in the time domain, filtering away the most rapid fluctuations.  This may not be a great issue since -- fortunately -- the gamma-ray induced Cherenkov light flashes in air last only a few nanoseconds, and thus the performance of Cherenkov telescopes cannot be made much worse, lest they would lose sensitivity to their primary task.  Still, since realistic electronics may reach resolutions on the order of 1\,ns, it would be desirable that the error budget should not have components in excess of such a value.

Among existing Cherenkov telescopes, this is satisfied by parabolic designs (e.g., MAGIC) but not by the Davies-Cotton concept (e.g., VERITAS or H.E.S.S.-I).  For example, in the H.E.S.S.-I telescopes the photons are spread over $\Delta$t\,$\sim$\,5 ns, with an rms width $\sim$\,1.4 ns (Bernl{\"o}hr et al.\ 2003).  For large telescopes, the time spread would become unacceptably large if a Davies-Cotton design were chosen, and those therefore normally are parabolic (e.g., MAGIC on La Palma; H.E.S.S.-II in Namibia, and MACE in Ladakh, India).  In principle, these then become isochronous -- apart from minute (sub-ns) effects caused by individual mirror facets being spherical rather than parabolic, or by the tesselated mirror facets being mounted somewhat staggered in depth.  Examples of distribution functions of the time spread in various designs of large Cherenkov telescopes are in Akhperjanian \& Sahakian (2004) and  Schliesser \& Mirzoyan (2005).

Also non-parabolic telescopes can be made effectively isochronous, if they have more than one optical element.  The two-mirror Schwarzschild-Couder design (Vassiliev et al.\ 2007) is attractive for smaller telescopes, not least because its smaller image scale permits smaller and less expensive focal-plane cameras.  Also, Schmidt-type telescopes may satisfy high demands on isochronicity, while also being compact, offering a wide field of view, and having a narrow point-spread function (Mirzoyan \& Andersen 2009).

\subsubsection{Focusing at `infinity'} 

The optical foci of Cherenkov telescopes are optimized to correspond to those heights in the atmosphere where most of the Cherenkov light originates, and the image of a distant star will then be slightly out of focus.  For a focal length of $f$=10~m, the focus shifts 1 cm between imaging at 10 km distance and at infinity, which for an $f$/1 telescope implies an additonal image spread of some cm.  In order to decrease the stellar image and not to take in too much of the night-sky background, it could be desirable (though not mandatory) to refocus the telescope on stars at `infinity'.   On some (especially larger) telescopes, such a possibility may be available anyhow since some refocusing can be required in response to mechanical deflections when pointing in different elevations, or as caused by nocturnal or seasonal temperature variations.  In the absence of such a possibility, a refocusing could still be achieved by placing a small optical lens in front of the detector.

\subsubsection{Placement of telescopes in an array}

The placement of telescopes in interferometers can be optimized for the best coverage of the $(u,v)-$plane (e.g., Boone 2001; Herrero 1971; Holdaway et al.\ 1999, Keto 1997; Mugnier et al.\ 1996; Thompson et al., 2001).  As the star gradually crosses the sky during a night, the projected baselines between pairs of telescopes change, depending on the angle under which the star is observed.  If the telescopes are placed in a regular geometric pattern, e.g., a repetitive square grid, the projected baselines are similar for many pairs of telescopes, and only a limited region of the $(u,v)-$plane is covered (on the other hand, redundant baselines result in better signal-to-noise for those).  Since stars rise in the east and move towards west, baselines between pairs of telescopes that are not oriented exactly east-west will trace out a wider variety of patterns.  Because of such considerations, existing amplitude interferometers (both optical and radio) locate their component telescopes in some optimal manner (e.g., in a Y-shape, or in logarithmic spirals, unless constrained by local geography). 

As concerns specifically the CTA, its smaller telescopes will be so numerous that, for most practical purposes, their exact placement should not be critical for interferometry -- a huge number of different baselines will be available anyway.  However, the situation is different for the very few large telescopes.  Avoiding placing them on a regular grid (such as a square) will offer a variety of baseline lengths, give a better coverage of the $(u,v)-$plane, and permit better image reconstruction. 

Possibly, not all telescopes in a complex such as CTA will be equipped, or be available for interferometry at any one time, and the issue then arises as to what subsets of telescopes preferentially should be selected for use.  Simulated observations with various such subsets are discussed by Dravins et al.\ (2012) and Jensen et al.\ (2010).

\subsubsection{Impact on observatory operations}

The impact of intensity interferometry on other Cherenkov array operations should not be significant.  One aspect is that -- while brighter moonlight may preclude accurate observations of the feeble atmospheric Cherenkov light -- measuring brighter stars in moonlight should be no problem for intensity interferometry, enabling efficient operations during both bright- and dark-Moon periods.  (Of course, all observations desire a minimum of background light, and at some point there might be issues if observing faint stars; however there are thousands of observable stars in the sky brighter than the moonlit sky background.) 

Potential sources for interferometry are distributed over large parts of the sky and permit vigorous observing programs from both northern and southern sites.  However, several among the hot and young stars belong to Gould's Belt, an approximately 30 million year old structure in the local Galaxy, sweeping across the constellations of Orion, Canis Major, Carina, Crux, Centaurus, and Scorpius, centered around right ascensions 5-7 hours, not far from the equator.  Thus, many primary targets are suitable to observe during northern-hemisphere winter or southern-hemisphere summer.  We note that this part of the sky is far away from the many gamma-ray sources near the center of the Galaxy (which is at right ascension 18 hours).

\subsection{Detectors and cameras} 

Typical Cherenkov telescopes have focal lengths on the order of 10\,m, providing a focal-plane image scale around 3 mm/arcmin.  A typical point-spread function of 3\,arcmin diameter thus corresponds to 1\,cm.  Detectors that are capable of photon counting with nanosecond time resolution include well-established vacuum-tube photomultipliers and large-size solid-state avalanche diode arrays that are under development.

A Cherenkov telescope typically holds several hundred photomultiplier tubes acting as `pixels' in its focal-plane camera.  The detectors and their ensuing electronics are naturally optimized for the triggering on, and the recording of, faint and brief transients of Cherenkov light and might not be readily adaptable for hour-long continuous recordings of bright stellar sources.  However, for intensity interferometry, only one pixel is required (at least in principle, though some provision for measuring the signal at zero baseline is required) and we note that in some telescopes (e.g., HEGRA and MAGIC; Lucarelli et al.\ 2008; O{\~n}a-Wilhelmi et al.\ 2004), the central camera pixel was specifically designed to be accessible for experiments without affecting any others.  Possibly, such central pixels could be usable to perform some experiments towards intensity interferometry as well.

However, even if a central pixel is accessible, it may not be possible to use it in its bare form.  If observing a bright source in broadband white light with a large telescope, the photon-count rate may become too large to handle, even for reduced photomultiplier voltages.  However, as discussed in Section \ref{5}, the signal-to-noise ratio in intensity interferometry is independent of the optical passband: the smaller photon flux in a narrow spectral segment is compensated by the increased optical coherence of the more monochromatic light.  This property can be exploited with some color filter to reduce the photon flux to a suitable level, or using a narrow-band filter tuned to some specific spectral feature of astrophysical significance.  For such uses, there should be some provision for some mechanical mounting in front of the detector to hold some small optical element(s).  A broader-band color filter could simply be placed immediately in front of a photomultiplier but a narrow-band filter could require additional arrangements.  Such filters are normally interferometric ones and those need to be used in collimated (parallel) light in order to provide a more precise narrow passband.  Since light reaching the Cherenkov camera is not collimated, some additional optics would then be required.  For non-collimated light, narrow passbands can still be realized with devices based on other principles, such as Christiansen filters which consist of an optical cell with crushed glass immersed in a liquid. At that wavelength where the indices of refraction for glass and liquid are equal, the cell is transparent, while at all other wavelengths, the light is reflected, scattered or refracted away at the many interfaces between the tiny glass pieces and the liquid (e.g., Balasubramanian et al.\ 1992).   

The further development and optimization of observational techniques is likely to involve experiments with other types of detectors, color filters, polarizers or other optical components which could be awkward to mechanically and electronically (re)place in the regular Cherenkov camera.  To minimize disturbances to the Cherenkov camera proper, it could be preferable to place an independent detector unit on the outside of its camera shutter lid.  Such constructions have already been made on existing Cherenkov telescopes, e.g. a 7-pixel unit on a H.E.S.S. telescope used a plane secondary mirror to put it into focus, and was used for experiments in very high time-resolution optical observations.  Its central pixel recorded the light curve of the target, while a ring of six ‘outer’ pixels monitored the sky background and acted as a veto system to reject atmospheric background events (Deil et al.\ 2008; 2009; Hinton et al.\ 2006).  For such devices, provision must also be made for electrical power supply and signal cables to/from the outside of these camera shutter lids.

\subsection{Signal handling} 

Electronic units, already used in photon-counting laboratory experiments preparing for stellar intensity interferometry, have time resolutions approaching 1\,ns, and the error budget should ideally not have components in excess of such a value (the signal-to-noise is proportional to the square root of the signal bandwidth; Section \ref{5}).  Telescopes may be separated by up to a kilometer or two, and the timing precision of the photon-pulse train from the detector to a central computing location should be assured to no worse than some nanosecond (for the timing of its leading pulse-edge; the pulse-width may be wider).  Such performance appears to be achievable by signal transmission in optical fibers (Rose et al.\ 2000; White et al.\ 2008).  Compared to metal cables, these have additional advantages of immunity to cross-talk and to electromagnetic interference, and also avoid the difficulty of maintaining a common ground and protection for the receiving electronics against (in some locations not uncommon) lightning strikes across the array.

\subsubsection{Correlators} 

A critical element of an intensity interferometer is the correlator which provides the averaged product of the intensity fluctuations $\langle\Delta I_1 \cdot \Delta I_2\rangle$ to be normalized by the average intensities $\langle I_1\rangle$ and $\langle I_2\rangle$ (Eq.\ \ref{intcorr4}).  The original interferometer at Narrabri used an analog correlator to multiply the photocurrents from its phototubes, and significant efforts were made to shield the signal cables from outside disturbances.  Current techniques, such as FPGA ({\it{Field Programmable Gate Arrays}}), permit to program electronic units into high-speed digital correlators with time resolutions of a few ns or better.   Such a correlator has been constructed at the University of Utah, digitizing the input signals at 200\,MHz with a 12-bit resolution.  To obtain the correlation, the samples are multiplied and summed up in an accumulation register.  

Similar units are also commercially available for primary applications in light scattering against laboratory specimens.  Such intensity-correlation spectroscopy is the temporal analog to the [spatial] intensity interferometry, and was developed after the subsequent theoretical understanding of intensity interferometry.  It was realized that high-speed photon correlation measurements were required and electronics initially developed in military laboratories were eventually commercialized, first by Malvern Instruments in the U.K. (Pike 1979), and nowadays offered by various commercial companies (e.g., Becker 2005).

At Lund Observatory, a series of digital correlators have been acquired over time from different commercial providers and used to pursue various experiments for high-speed photon counting in optical astrophysics, including studies of atmospheric scintillation at the observatory on La Palma (e.g., Dravins et al.\ 1997), and in searches for high-speed astrophysical phenomena, when connected to the {\it{OPTIMA}} photometric instrument of the Max-Planck-Institute of Extraterrestrial Physics (Kanbach et al.\ 2008).   While the early correlators were impressively voluminous rack-mounted units, their electronics have since been miniaturized and current units are very small and easily transportable items, built around FPGAs, accepting many input channels, running at sampling frequencies up to 700\,MHz, handling continuous photon-count rates of more than 100\,MHz per channel without any deadtimes, with on-line data transfer to a host computer.  Their output contains the cross correlation function between the two telescopes (as well as autocorrelation functions for each of them), made up of about a thousand points.  For small delays (where most of the intensity interferometry signal resides), the sampling of the correlation functions is made with the smallest timesteps, which increase in a geometric progression to large values to reveal the full function up to long delays of seconds and even minutes.  Individual photon events are normally not saved, although that is possible for moderate count rates below about 1\,MHz.  It is believed that their electronic performance is now adequate for full intensity interferometry experiments.

Still, the use of such correlators is not without issues.  For realistic observations of bright objects, the searched-for signal is only a tiny fraction of the full (Poisson-noise) intensity correlation in the raw data, and the signal must be analyzed with many bits of resolution in the digital case, and with a high degree of shielding in the analog case.   
An advantage with firmware correlators is that they produce correlation functions in real time, processing very large amounts of photon-count data, and eliminating the need for their further handling and storage (e.g., existing correlators, using 10 input channels, each running at 50\,MHz during one 8-hour observing night process more than 10\,TB of photon-count data).  A disadvantage is that, if something needs to be checked afterwards, the full set of original data is no longer available, and alternative signal processing cannot be applied.

An alternative approach (at least for limited photon-count rates) is to time-tag each photon count and store all data, and then later perform the correlation analyses off-line.  The data streams from multiple telescopes can then be cross-correlated using a software correlation algorithm, permitting the application of also digital filtering to eliminate possible interference noise from known sources, and also to compute other spatio-temporal parameters, such as higher-order correlations between three telescopes or more, which in principle may contain additional information.  A disadvantage is that this requires a massive computing effort and it is not clear whether it realistically permits much more than standard correlations to be computed; also possible observational problems may not get detected while observations are in progress but only at some later time.  Such a capability was foreseen in the design study for {\it{QuantEYE}}, a proposed very high-time resolution instrument for extremely large telescopes (Dravins et al.\ 2005, 2006), and verified in the construction and operation of the {\it{AquEYE}} and {\it{IquEYE}} instruments, the latter used also at the European Southern Observatory in Chile (Naletto et al.\ 2007; 2009; 2010).

\subsection{Delay units} 

Besides the correlator, another piece of electronics is required for real-time intensity interferometry, namely to implement a continuously variable time delay that compensates for the relative timing of the wavefront at the different telescopes, as the source moves across the sky (Eq.\ \ref{rotationsynth}).  Various solutions are possible for either analog or digital signal handling: for example, one test unit constructed at Lund Observatory comprises a continuously variable and programmable delay of up to a few {$\mu$}s of the photon-pulse train (corresponding to differential light travel distances of maybe half a km), using a buffer memory into which the photon pulses are read in and read out almost simultaneously, but with a programmable and continuously changing readout frequency, thus slightly stretching or squeezing the electronic pulse-train to create the required and continuously changing delay.     

If such a delay unit is not used, the maximum correlation signal in a multichannel digital correlator will appear not in the channel for zero time delay between any pair of telescopes, but rather at that channel which corresponds to a delay equal to the light-time difference between telescopes along the line of sight towards the source.  This arrangement is feasible already with existing digital correlators since these can be programmed to measure the correlation at full time resolution also at time coordinates away from zero (though it could require frequent readouts since these delays continuously change as the star moves overhead).

However, in case each photon count has been time-tagged and stored, such arrangements are not required since the delays can be introduced by software in the later off-line data analysis.

\begin{figure}
\centering
\includegraphics[width=6cm, angle=180]{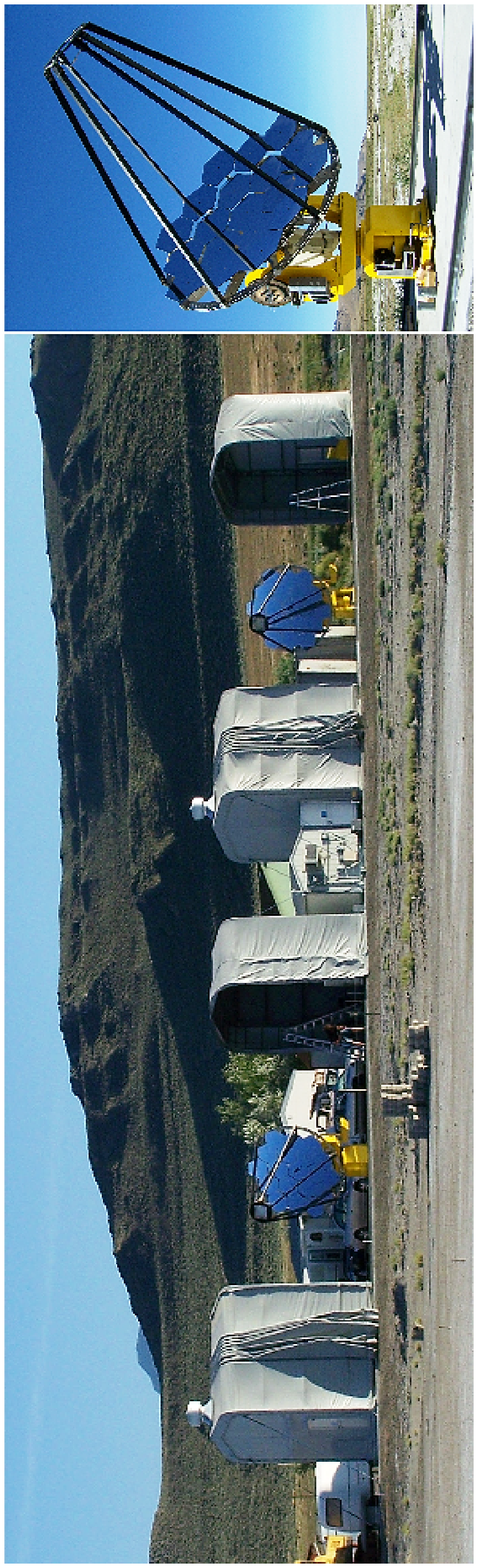}
\caption{The StarBase 3\,m telescopes are protected by buildings which can be rolled open for observation (left) The control room is located in a smaller building located between the two telescopes. Right: Close-up view of one telescope.  \label{starbase} }
\end{figure}

\section{Experimental work} 

As preparatory steps towards realizing full-scale stellar intensity interferometry, different laboratory and field experiments are being carried out at various institutes.

\subsection{The StarBase telescopes} 

Even if existing Cherenkov telescope facilities have been supportive in giving access to their telescopes for various verifications and tests, any more extensive experimental work will be easier carried out at a facility where the instrumentation can be modified without having to remain compatible with Cherenkov observations during the following or even the same night.

For such purposes, a testbed observatory has been set up at the site of a geothermal diving facility (Bonneville Seabase 2012) in Grantsville, some 60\,km west from Salt Lake City, Utah.  This {\it{StarBase}} (2012) is equipped with two air Cherenkov telescopes on a 23\,m east-west baseline (Figure \ref{starbase}; LeBohec et al.\ 2008b; 2010).  Those telescopes had earlier been used in the Telescope Array experiment (Aiso et al.\ 1997) operated until 1998 on the Dugway proving range.  Each telescope is a 3\,m diameter, $f/1$ light collector of the Davies-Cotton type, composed of 19 hexagonal mirror facets $\sim$\,60\,cm across. This design is typical for air Cherenkov telescopes and secondary optics tested on these telescopes may be directly used on others for larger-scale tests. The telescope mounts are alt-azimuthal with the motion around both axes controlled by tangential screws and absolute encoders with a few arcsecond resolution. The tracking model parameters are being optimized but the absolute pointing accuracy is better than four arcminutes and can be compensated by online corrections. The optical point spread function (PSF) with full width at half maximum $\sim$\,6\,arcmin is dominated by spherical aberration of individual mirror facets.  This, however, is untypical of large Davies-Cotton Cherenkov telescopes where it typically is $\sim$\,3\,arcmin. This difference is due to the facets of the StarBase telescope being much larger in proportion to the telescope diameter than usual. For example, the ones at VERITAS (2012) are 12\,m diameter $f/1$ light collectors with 350 mirror facets $\sim$\,60\,cm across.  Interestingly, this lower angular performance of the StarBase light collectors make them suitable for larger-scale implementations since the PSF linear extent is very comparable to that in large telescopes such as in VERITAS and the aperture ratio is the same.

Using conservative parameters for the StarBase telescopes, it is estimated that a 5 standard-deviation measurement of a degree of coherence $|\gamma(r)|^2=0.5$  will require an observation time of one hour for a star of m$_{V}$\,=1, and some 6 hours for m$_{V}$\,=\,2 (for $|\gamma(r)|^2\sim\,1$, these times should be divided by four).  Thus, the facility is suitable for observing bright stars, e.g., to measure the degree of coherence for unresolved objects.  The distance between the telescopes being 23~m (with smaller projected baselines when observing towards the east or the west), at $\lambda$\,=400\,nm, such stars have to be below $\sim$~3~mas in diameter, and an unresolved star suitable for calibration should be less than $\sim$~1~mas.  Several good candidates are available, e.g., $\alpha$~Leo, $\gamma$~Ori, $\beta$~Tau or $\eta$~UMa.  The facility also permits to search for coherence modulation resulting from orbital motion in the binary Spica with an $a$\,=\,1.5\,mas semi-major axis or possibly even Algol ($a$\,=\,2.2\,mas, m$_{V}$=2.1).

\subsection{StarBase cameras} 

For the StarBase telescopes, cameras with control electronics are being constructed, for either off- or online analysis of the data.  The cameras must be suitable for intensity interferometry also with one single telescope (to provide the zero-baseline correlation), and thus provide two channels.  The telescopes are made compatible with an independent Cherenkov camera in the focal plane, and the intensity interferometry units are mounted on the outside of its camera lid.  This is achieved by using a large enough mirror making a $45^\circ$ angle with the telescope optical axis so all secondary optics is parallel to the focal plane.

The camera optics must be able to select a narrow optical passband and concentrate the light on one or two photodetectors if the zero baseline correlation is to be measured; the latter by using a beamsplitter.  As already mentioned, narrow optical passbands may be required both to moderate the flux from bright stars, and for selecting astrophysically interesting spectral features.  Also, replacing the beamsplitter by a dichroic mirror would allow simultaneous measurements in two optical passbands.

\subsection{Laboratory experiments} 

Various laboratory experiments simulating aspects of intensity interferometry are carried out at different institutes.  For example, at Lund Observatory, an intensity interferometer has been set up in an optics laboratory, simulating observations of a star with two telescopes, employing high-speed photon-counting detectors with real-time digital cross correlation of their intensity-fluctuation signals, concluding with a determination of the angular extent of the source.  The purpose is to verify and develop some of the techniques required for future full-scale observations, and to better understand issues such as effects from partially polarized light, detector imperfections (e.g., afterpulsing), and data handling.

\begin{figure}
\centering
\includegraphics[width=11cm]{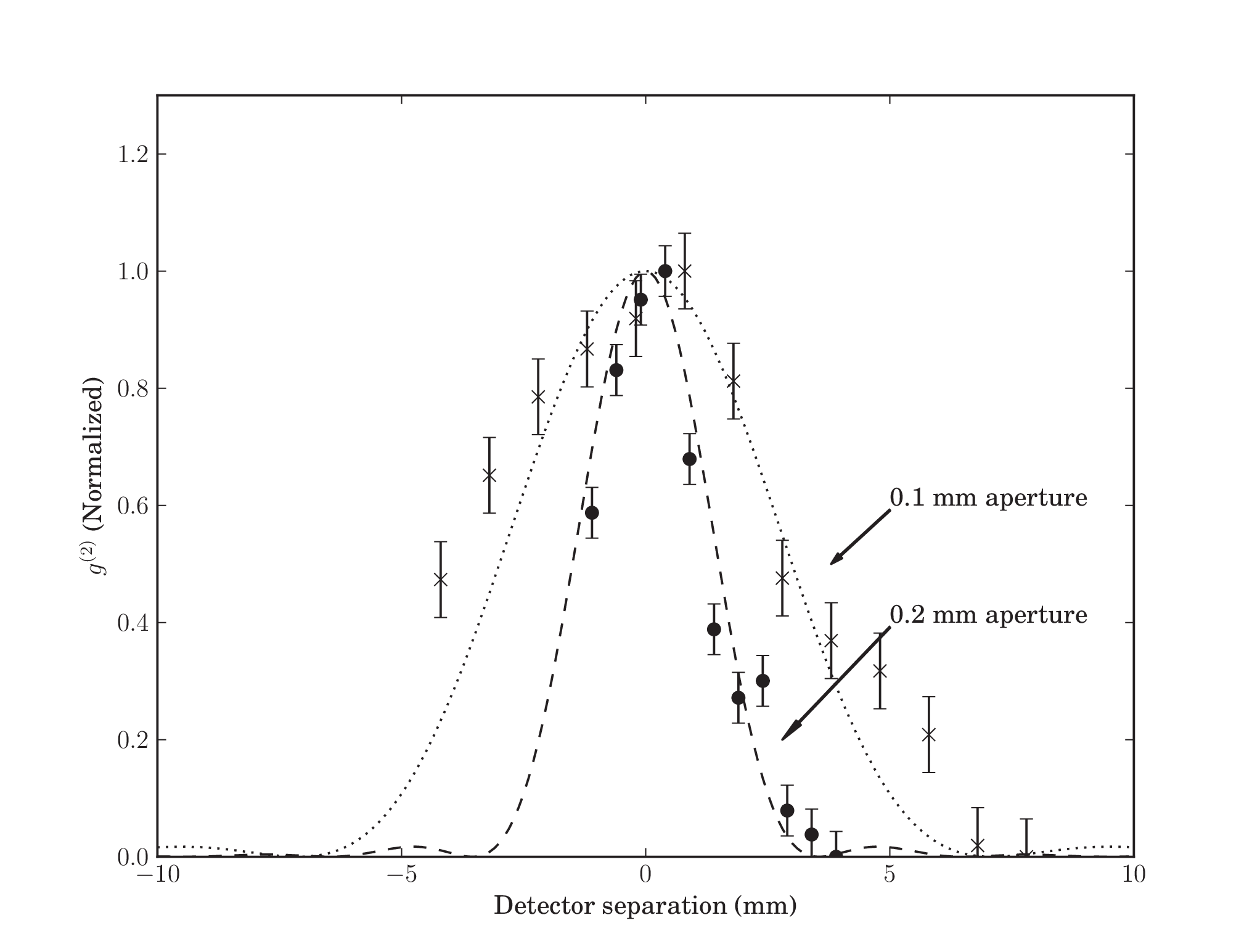} 
\caption{Results from a laboratory experiment, simulating observations of differently large stars with a two-telescope intensity interferometer.  The plot shows the measured (normalized) second-order coherence for two artificial `stars', being illuminated pinholes of different diameters.  Each measured point results from 400\,s of integration; error bars indicate experimental reproducibility while the dashed and dotted curves show the theoretically expected runs of the coherence functions.}
\label{labexperiment}
\end{figure}

An artificial star is provided by a small illuminated pinhole, while the telescopes are refractors constructed as optical-bench units, whose separation is varied by translating them on optical carriers perpendicular to the observing direction.  Light is focused onto single-photon-counting avalanche photodiodes (SPADs), enabling photon-count rates up to some 10\,MHz.  The pulse-train output (electronic TTL standard) is fed to a hardware correlator for real-time cross correlation of the data streams from both telescopes.  The measured intensity correlation is normalized to unity for zero baseline, and examples for  differently sized `stars' are in Figure \ref{labexperiment}.

However, the realization of these experiments was preceded by several less successful attempts.  In particular, as discussed in Section \ref{5}, intensity interferometry is primarily sensitive to sources of high brightness temperature but limited in observations of cool ones, and of course exactly the same conditions apply to any laboratory setup as for stars in the sky.  The source must be small enough to produce an extended diffraction pattern that can be sampled by the interferometer, and be bright enough to produce acceptable photon count rates.  While there are many stars in the sky with T$_{eff}$=\,10,000\,K or more, to produce a correspondingly brilliant laboratory source is much more challenging.  It should be recalled that the method of intensity interferometry implicitly assumes that the light is chaotic (with a Gaussian amplitude distribution; Bachor \& Ralph 2004; Foellmi 2009; Loudon 2000; Shih 2011), i.e., the light waves undergo random phase shifts so that an intensity fluctuation results, which then bears a simple relation to the ordinary first-order coherence.  While this must be closely satisfied for any thermal source, it is not the case for a laser which, ideally, never undergoes any intensity fluctuations anywhere.  The spatial extent of a laser source therefore cannot be measured by intensity interferometry, and a laser is not an option to enhance the brightness of such an artificial star.

In initial attempts to achieve a high surface brightness for the illuminated pinhole serving as the `star', the very small emission volume of a high-pressure Hg arc lamp was focused onto it, and a narrow-band optical filter singled out its brightest emission line ($\lambda$\,546 nm).   Although this represents about the highest black-body brightness temperature (some 3,000\,K) that can readily be obtained with ordinary laboratory equipment for a non-laser source, the photon-count rates still turned out to be slightly too low for measurements with conveniently short integration times.  Since the signal-to-noise ratio increases with the number of photons per unit frequency bandwidth but does not depend on the optical passband, arrangements were made to instead obtain highly intense quasi-monochromatic sources.  Line profiles from several emission-line lamps of various atomic species were measured with a Fourier transforms spectrometer of extremely high resolution to identify those that produced the brightest and narrowest emission lines, and that also were sufficiently isolated within their spectra to be selectable by narrow-band interference filters.  As the best among these, a Na I lamp was chosen, somewhat improving the signal which, however, still remained marginal.  As the final choice, quasi-monochromatic chaotic light was produced by scattering monochromatic He-Ne laser light against microscopic (0.2\,$\mu$m diameter) polystyrene spheres, suspended in a cm-sized cuvette with room-temperature water.  These microspheres undergo thermal (Brownian) motion, producing a slightly Doppler-broadened spectral line which is extremely narrow (order of kHz), and was estimated to have an effective brightness temperature around T$_{eff}$=\,60,000\,K, permitting meaningful measurements already with minute-long integration times.  Such scattered laser-light is used for various laboratory photon-correlation measurements of time variability (Becker 2005), but we are not aware of any previous such experiment with a spatial intensity interferometer.

\subsection{Full-scale observations with VERITAS} 

As the first full-scale test toward implementing intensity interferometry with Cherenkov telescope arrays, pairs of the 12\,m telescopes in the VERITAS array at the Fred Lawrence Whipple Observatory in Arizona were used to observe a number of stars, with pairs of its telescopes interconnected through digital correlators.  Baselines between pairs of its four telescopes then ranged between 34 and 109\,m (one of the telescopes has since been moved).  

For these observations, starlight was detected by a photon-counting photomultiplier in the central pixel of the regular Cherenkov-light camera, the outgoing photon pulses were digitized using a discriminator, then pulse-shaped and transmitted from each telescope via an optical cable to the control building where they entered a real-time digital cross correlator, computing the cross correlation function for various time delays.  Continuous count rates up to some 30\,MHz were handled, limited by the digitization and signal-shaping electronics.  These experiments were not intended to measure astrophysical quantities but to gain experience in operating observations with a full-scale observatory.  Nevertheless, we believe these experiments represent the first case of optical astronomical telescopes having been connected for real-time observations through $e$-interferometry by digital software rather than by optical links (in some sense following in the footsteps of radio $e$-VLBI).  For details, see Dravins \& LeBohec (2008).

\section{Further possibilities} 

The technique of intensity interferometry may be used not only with arrays of Cherenkov telescopes, and not only for obtaining source images.  In this Section, we point out a few other potential applications.

\subsection{Extremely large telescopes} 

One of the goals for extremely large optical telescopes (ELTs) with apertures in the 30--40 m range, is diffraction-limited imaging using adaptive optics, expected to initially become feasible at longer wavelengths in the near-infrared.  Although the resolution offered is rather coarser than with the long baselines in Cherenkov telescope arrays, also ELTs offer possibilities for intensity interferometry, provided they are outfitted with a suitable high-speed photon-counting instrument.  This potential was analyzed in the design study of the {\it{QuantEYE}} instrument (Barbieri et al.\ 2007; Dravins et al.\ 2005; 2006).  There, the ELT entrance pupil was optically sliced into a hundred segments, each feeding a separate photon-counting detector.  Different means of electronically combining the signal in software would yield either a photometric signal of very high time resolution using the collecting area of the entire telescope, or -- by suitable cross correlations -- realize intensity interferometry between various pairs of telescope subapertures.  Being immune against atmospheric turbulence, such observations could be made when seeing conditions are inadequate for adaptive optics, and would be practical already with the main mirror being only partially or sparsely filled with mirror segments (a situation likely to last for several years during any ELT construction phase, given the huge number of segments that make up the primary).  Since intensity interferometry has no limitations at short wavelengths other than the atmospheric cutoff, the achievable spatial resolution will be superior to that feasible by infrared adaptive optics by a factor of 2 or 3 (besides viewing astrophysically different emission from the source at much shorter wavelengths).  These concepts towards such an instrument for ELTs have been further developed in the construction and operation of the {\it{AquEYE}} and {\it{IquEYE}} instruments, used at the Asiago and ESO La Silla observatories (Naletto et al.\ 2007; 2009; 2010).

Although mirror segments on ELTs are much smaller than Cherenkov telescopes, they offer certain advantages: their image quality is arcseconds or better, which essentially eliminates background light from the night sky, and in particular permits the use of small detectors of very high quantum efficiency, such as single-photon-counting avalanche diodes, which are as yet not fully adapted to the large optical point-spread functions of Cherenkov telescopes.  The high degree of optical collimation permits the use of interference filters with very narrow bandpass to isolate spectral lines, and since the optical paths are isochronous, there is no optical limitation in how fast electronics that can be used.  Thus, also extremely fast detectors could be utilized (e.g., Margaryan 2011) to improve the signal-to-noise ratio and reach fainter sources, perhaps even extragalactic ones.  Although the finite size of the ELT aperture limits the extent of the $(u,v)-$plane covered, this can be sampled very densely, and an enormous number of baseline pairs can be synthesized, assuring a complete sampling of the source image, and its stable reconstruction.

\subsection{Higher-order spatio-temporal correlations} 

The quantum theory of optical coherence (e.g., Glauber 1963abc; 2007; Mandel \& Wolf 1995) describes how one can define correlations between arbitrarily many spatial and/or temporal coordinates in the volume of light (`photon gas') being received from a source.  The spatial intensity interferometer is only one special case of such more general spatio-temporal correlations, in that it measures the cross correlation between the intensities at {\it{two}} spatial locations, at {\it{one}} instant in time.   

However, using telescope arrays, and given that their photon detectors provide data streams which can be analyzed at will, one can construct, e.g., third-order intensity correlations, $g^{(3)}$, for a system of three telescopes: $\langle I(r_1, t_1) I(r_2, t_2) I(r_3, t_3) \rangle$, where the temporal coordinates do not necessarily have to be equal.  In principle, such and other higher-order spatio-temporal correlations in light may carry additional information about the source from where the light has been emitted and thus is of relevance for astronomy where information about the source has to be extracted from more or less subtle properties of its radiation received (Jain \& Ralston 2008; Ofir \& Ribak 2006a).

Although, in the recording of higher-order correlations, also the relative noise level increases (possibly demanding very large telescopes for certain measurements; Dravins 1994), all sorts of higher-order correlations can in principle be obtained without any additional observational effort if the digital signals from each telescope are avaible for further manipulation in either hard- or software.  Thus, one could calculate correlations among also all possible triplets and quadruplets of telescopes, possibly enabling a more robust full reconstruction of the source image (Ebstein 1991; Fontana 1983; Hyland 2005; Marathay et al.\ 1994; Sato et al.\ 1978; 1981; Schulz \& Gupta 1998; Zhilyaev 2008).

A different special case is a `temporal' intensity interferometer, measuring the cross correlation between the intensities at {\it{one}} spatial location, but at {\it{two}} or more instants in time.  The information obtained is then not that of the spatial coherence (i.e., of the source's spatial extent), but rather of its temporal coherence, i.e., its spectral extent.  Analogous to the spatial information extracted from intensity interferometry, this photon-correlation spectroscopy directly provides the spectral extent of the source with respect to the temporal `baseline' over which it has been observed.  Since this temporal delay can be quite large (1\,ms corresponds to 1\,kHz resolution in the electromagnetic spectrum), the spectral resolution obtainable can be enormously much higher than feasible with conventional spectrometers.  This has laboratory applications in light-scattering experiments and in astronomy appears to be required for spectrally resolving the emission components from natural lasers operating in very luminous sources such as $\eta$~Carinae (Johansson \& Letokhov 2004, 2007; Letokhov \& Johansson 2009).  These are theoretically expected to be extremely narrow: on order 100\,MHz (demanding spectral resolution $\lambda$/$\Delta\lambda\sim10^8$, and measurements over temporal delays of 10\,ns).   The prospects of resolving such emission with intensity correlation measurements were discussed by Dravins (2008), Dravins \& German{\`{a}} (2008), and Johansson \& Letokhov (2005).  Also here, correlations of higher order than two may convey additional information (Gamo 1963).

\section{Outlook} 

Looking back in time, the idea of using interferometry to measure stellar diameters appears to have been first suggested by Fizeau (1868), and carried out by St{\'e}phan (1874) who placed a two-aperture mask over a 80 cm reflector at Marseille Observatory, but realized that stars could not be resolved over this short baseline.  In the 1920's, Michelson \& Pease (1921) operated a 6-meter interferometer mounted on the 100-inch Hooker telescope on Mt.Wilson, and succeeded in measuring diameters of a few giant stars, while their later 15-meter instrument proved mechanically too unstable for practical use (Hariharan 1985).  

The demanding requirement to maintain stable optical path differences during observations to a fraction of an optical wavelength caused the technique to lay dormant for half a century, until Labeyrie (1975) succeeded in measuring interference fringes between two separated telescopes.  This success triggered the construction of a whole generation of optical amplitude interferometers which have now provided tantalizing glimpses of the individualities among our neighboring stars.

These breakthroughs in amplitude interferometry during the 1970s are said to have been the specific reason why the plans to build a successor to the original Narrabri intensity interferometer (designed around that very time) were not realized, and (as far as astronomy is concerned), the technique has now been dormant for several decades.  However, recent progress in instrumentation and computing technology has been extraordinary.  High-speed photon-counting detectors and hardware correlators are commercially available, and new mathematical algorithms allow for image reconstruction. The most valuable components -- large flux collectors -- exist in the form of air Cherenkov telescopes, with many more coming.  All of this has sparked a renewed interest in astronomical intensity interferometry, and a first workshop (since very many years) on stellar intensity interferometry was held not long ago (LeBohec 2009).  For Cherenkov telescope arrays, the ongoing upgrade of the VERITAS array in Arizona includes provisions for intensity interferometry, as does the preparatory phase of the planned international Cherenkov Telescope Array.  Thus, long after the pioneering experiments by Hanbury Brown and Twiss, the technological developments carry the promise of achieving a basic but difficult goal: to finally be able to view our neighboring stars as the extended and most probably very fascinating objects that they really are.

\section{Acknowledgements}

The work at Lund Observatory is supported by the Swedish Research Council and The Royal Physiographic Society in Lund.  Stephan LeBohec acknowledges support from grants SGER \#0808636 of the National Science Foundation.  For our field experiments, we thank the VERITAS collaboration for providing access to its telescope array.  During those observations, valuable support and advice was provided by Michael Daniel (then at University of Leeds, now Durham University).  We are grateful to Linda Nelson and George Sanders for hosting the StarBase observatory on the site of the Bonneville Seabase diving resort in Utah.  Work at the StarBase facility involves several persons from the University of Utah, including Ben Adams, Derrick Kress, Edward Munford, Ryan Price, Janvida Rou, Harold Simpson, and Jeremy Smith.   Early experiments in laboratory intensity interferometry at Lund Observatory involved also Ricky Nilsson and Helena Uthas, where we also acknowledge the technical advice by Nels Hansson, Bo Nilsson, and Torbj{\"{o}}rn Wiesel.  Discussions and site visits about the possible use of the MAGIC Cherenkov telescopes involved several people, including Juan Abel Barrio and Fabrizio Lucarelli (Madrid), Thomas Schweizer, the late Florian Goebel (Munich), Juan Cortina (Barcelona), and Markus Gaug (La Palma).  In the context of evaluating the use of the planned Cherenkov Telescope Array for intensity interferometry, valuable comments were received from several colleagues, including Michael Andersen, Cesare Barbieri, Isobel Bond, Stella Bradbury, Hugues Castarede, Michael Daniel, the late Okkie de Jager, Willem-Jan de Wit, Wilfried Domainko, C{\'{e}}dric Foellmi, Werner Hofmann, Richard Holmes, David Kieda, Antoine Labeyrie, Lennart Lindegren, Hans-G{\"{u}}nter Ludwig, Giampiero Naletto, Aviv Ofir, Guy Perrin, Andreas Quirrenbach, Erez Ribak, Stephen Ridgway, Joachim Rose, Diego F.\ Torres, Gerard van Belle, and Luca Zampieri.  We acknowledge constructive comments by an anonymous referee, and a suggestion by journal editor Phil Charles about including the subsection on observing novae.  The part of the work connected with the CTA received funding from the European Union's Seventh Framework Programme ([FP7/2007-2013] [FP7/2007-2011]) under grant agreement nr.\ 262053.

\end{document}